\documentclass[12pt,preprint]{aastex}
\usepackage{color}
\usepackage{epsfig}
\newcommand{\vdag}{(v)^\dagger}
\defcitealias{jeffery93a}{JH93}

\slugcomment{Draft version July 2014.}

\shorttitle{Chemical composition of DY\,Cen}
\shortauthors{Pandey \& ...}

\begin{document}

\title{On the binary helium star DY\,Centauri: Chemical composition and evolutionary state}

\author{Gajendra Pandey}
\affil{Indian Institute of Astrophysics;
Bangalore,  560034 India}
\email{pandey@iiap.res.in}
\and
\author{N.\ Kameswara Rao}
\affil{Indian Institute of Astrophysics;
Bangalore,  560034 India}
\email{nkrao@iiap.res.in}
\and
\author{C.\ Simon Jeffery}
\affil{Armagh Observatory; Collage Hill, Armagh BT61 9DG, N. Ireland}
\email{csj@arm.ac.uk}
\and
\author{David L.\ Lambert}
\affil{The W.J. McDonald Observatory and Department of Astronomy, University of Texas at Austin; Austin,
TX 78712-1083}
\email{dll@astro.as.utexas.edu}

\begin{abstract}

DY\,Cen has shown a steady fading of its visual light
by about 1 magnitude in the last 40 years suggesting a secular increase in its effective temperature.
We have  conducted  non-LTE and LTE abundance analyses to determine the star's effective temperature,
surface gravity, and chemical composition using high-resolution spectra obtained over two decades.
The derived stellar parameters for three epochs suggest that DY\,Cen has evolved at a constant
luminosity and has become hotter by about 5000 K in 23 years.
We show that the derived abundances remain unchanged for the three epochs. The derived abundances
of the key elements, including F and Ne, are as observed for the extreme helium stars resulting
from a merger of an He white dwarf with a C-O white dwarf. Thus, DY\,Cen by chemical composition appears
to be also a product of  a merger of two white dwarfs. This appearance seems to be at odds with the 
recent suggestion that DY\,Cen is a single-lined spectroscopic binary.

\end{abstract}

\clearpage

\section{Introduction}

The hydrogen-deficient giant DY\,Centauri is commonly
known as an R Coronae Borealis (RCB) variable
\citep{hoffleit30}, although no RCB-type activity has been observed since that reported in 1930.
From its colour, it was known to be significantly hotter than most other RCB stars \citep{kilkenny84b},
whilst a high-resolution spectrum obtained in 1987 showed it to have an effective temperature and
surface composition similar to that of the hotter extreme helium (EHe) stars \citep{jeffery93a}
(\citetalias{jeffery93a}), albeit with an apparently low iron abundance and a high (10\%) hydrogen abundance. An
identification of RCB and EHe stars with  stars evolving on a post white dwarf merger track has become
increasingly strong in recent years \citep{saio02,pandey06a,clayton07,jeffery11a}; the assumption
was that DY\,Cen also lies on this track.

Recent evidence suggests that this may not be so.
The conclusion that DY\,Cen represents a substantially different type of star to the RCB and EHe stars
was hinted at by \citet{demarco02}, who highlighted a number of discrepant radial-velocity
measurements, and demonstrated a systematic increase in V magnitude during an interval of some 80
years. \citet{rao12} demonstrated that DY\,Cen is a single-lined spectroscopic binary with an
orbital period of 39 days; all other EHe and RCB stars are single stars, as expected for a
white-dwarf-merger origin. DY\,Cen and another ``RCB" star with significant hydrogen, V854\,Cen,
show other differences to normal RCB stars. {\it Spitzer Space Telescope} observations show C$_{60}$ in
both DY\,Cen and V854\,Cen, but not in other RCB stars \citep{garcia11c}. 
DY Cen's spectrum is also unusual in that nebular emission lines
are present at maximum light. Note that, there are three other `hot' RCB stars
with emission lines: MV\,Sgr, V348\,Sgr and HV\,2671. DY\,Cen and MV\,Sgr seem to have RCB abundances,
while V348\,Sgr and HV\,2671 do not (see \citet{clayton2011}, and references therein).
The present paper was stimulated in part by the systematic and secular fading of DY\,Cen
from a visual magnitude of 12.2 in 1970 to nearer 13.2 in 2010. To explain
this requires either a steady increase in extinction, a fall in intrinsic brightness, a change in
effective temperature (which changes the bolometric correction) or a combination of all three. Given
the magnitude and rapidity of the change, any one of these has profound consequences for
interpreting the evolutionary status of DY\,Cen.

This paper 
first
examines the implications of the visual-magnitude variation assuming evolution at constant
luminosity. It then carries out a fine-analysis for effective temperature, surface gravity and
chemical composition using high-quality spectra obtained over two decades. 
Judged by chemical composition, especially by the F and Ne abundances, DY\,Cen closely resembles
hot EHes. Since EHes are generally considered to be single stars and formed by the merger of a 
Helium white dwarf with a C-O white dwarf, identification of DY\,Cen with the EHes challenges the 
recent identification of DY\,Cen as a single-lined spectroscopic binary. 

\section{The secular fading of DY\,Cen}

\citet{drilling86b}  noted that, if the EHe and RCB stars share the same luminosity, they  show
a steady fading of the absolute visual magnitude towards higher effective temperature ($T_{\rm eff}$). 
The photometric variation of DY\,Cen with time was summarized  by 
\citet{demarco02} and \citet{rao13} showing a 
fading in visual magnitude by about 1.3 magnitudes over the previous century,
with dramatic change in the nebular emission line fluxes,
prompting the suggestion that it could be associated with a 
secular increase in effective temperature on a timescale of decades.  
%

 In order to calibrate the star's fading, 
we have taken the theoretical spectral energy 
distributions (SED) for a grid of line-blanketed hydrogen-deficient model atmospheres 
\citep{behara06}, computed with a chemical composition appropriate to that  of DY\,Cen.
Since the SED total flux is proportional to the fourth power of $T_{\rm eff}$, 
each SED in the model grid was  divided by this quantity in order to 
apply a  constant luminosity approximation. 
Each SED was then convolved with the  $B$- and $V$-band filter response functions 
given in Table 2 of \citet{bessell90}, integrated to obtain the total flux in the $V$-band, 
and converted to a magnitude. 

An arbitrary zero-point is obtained by comparing the theoretical $V$-magnitude corresponding to
$T_{\rm eff}=19\,500$K in 1987 May
 measured spectroscopically by \citetalias{jeffery93a} with a mean value of $V=12.784$ 
measured by \citet{pollacco91a} in 1987 May-June. 
The zero-point corrected theoretical $V$ magnitudes and the B--V
colours are shown as a function of $T_{\rm eff}$ in Figure 1, together with observed
$V$ magnitudes from 1930, 1970, 1987 \citep{hoffleit30,marino71, pollacco91a} and 2007
(AAVSO\footnote{http://www.aavso.org}). Hoffleit's photographic magnitudes are converted to $V$
magnitude by a recipe given by \citep{demarco02}.

\begin{figure}
\epsscale{1.00}
\plotone{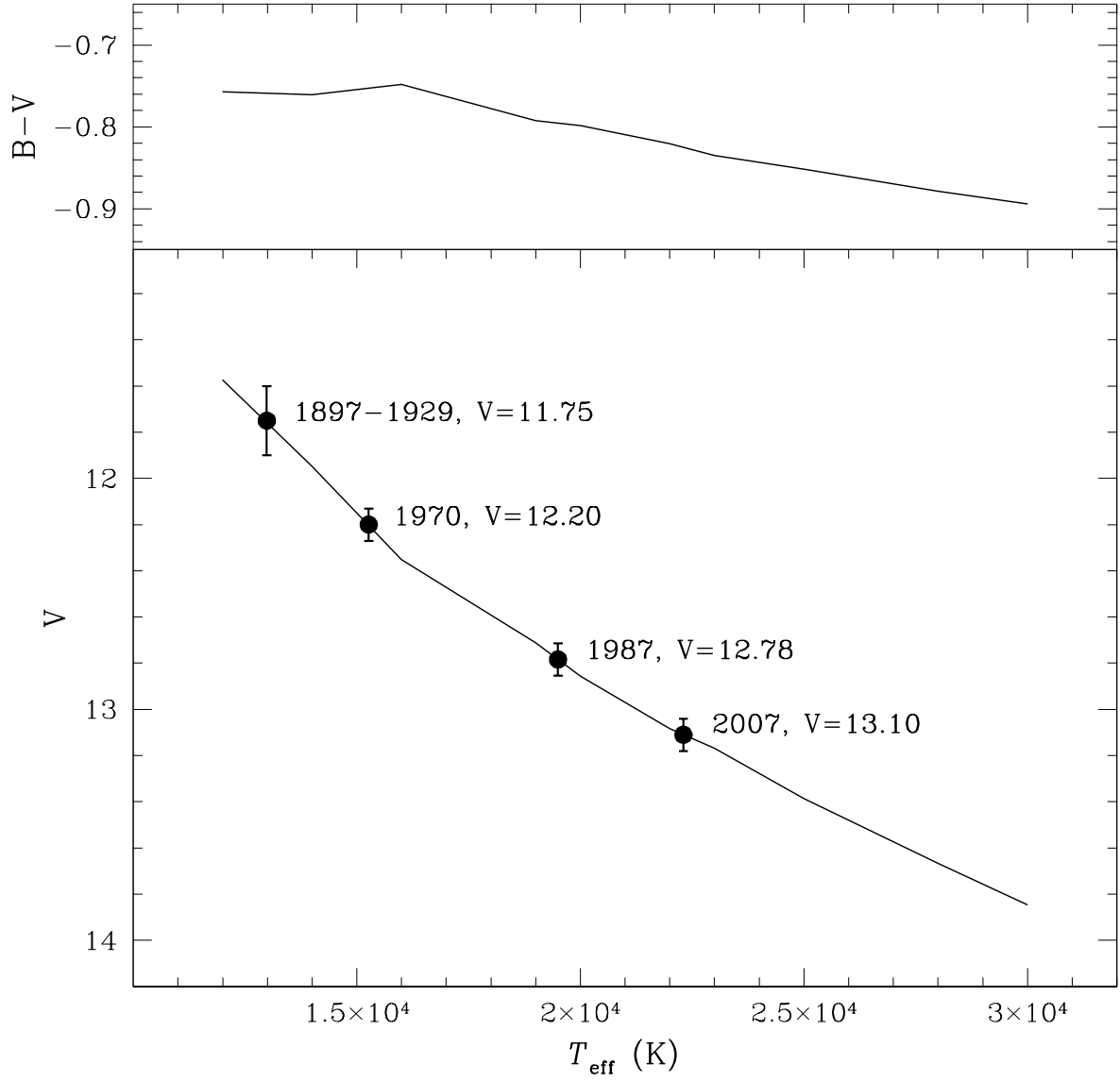}
\caption{Lower panel: Theoretical zero-point corrected V magnitudes versus $T_{\rm eff}$.
Observed $V$  magnitudes are shown by filled circles, with errors and epoch. The
effective temperatures for the 1987 point is taken from spectroscopy.
Upper panel: Theoretical B$-$V colors versus $T_{\rm eff}$.
\label{f:S3mags}}
\end{figure}

Figure\,\ref{f:S3mags} suggests that, if the luminosity has remained constant, an explanation for
the secular fading of DY\,Cen can be provided by an increase in effective temperature from 
about 14\,000 K in 1930, to about 25\,000 K in 2010. It remains to be shown that this 
is the {\it only} explanation
which can account for the observations. 

In the following sections, high-resolution optical spectra 
from 1987 to 2010 are analysed to
determine if the star has evolved at constant luminosity with concomitant variations in
effective temperature and surface gravity. In addition, the chemical composition of DY Cen
is determined anew using the superior spectra now available.

\section{Observations}

High-resolution spectra of DY\,Cen, are available from the nights of 1987  April 17,  
2002  June 24 and several nights between  2010 February 27 to 2010 March 2. 
All spectra were obtained at maximum light, there having been no minima recorded over the period
of these observations according to visual estimates from AAVSO. 
The data were reduced using standard procedures appropriate to the instrument, as described by \citetalias{jeffery93a} and \cite{rao12}.

The 1987 spectrum was obtained with CASPEC, the Cassegrain \'echelle spectrograph 
on the 3.6m telescope of the European Southern Observatory at  La Silla, Chile. The spectrum
covers the spectral range 4000 -- 4900\AA\ at a (2-pixel) resolution of about
22\,000.
The spectrum was analyzed in detail by \citetalias{jeffery93a}.

The 2002 spectrum was obtained with UCLES, the University College \'echelle spectrograph on the
3.9m Anglo-Australian Telescope at Siding Spring Observatory, Australia. The spectrum
covers the spectral range 3800 -- 5100\AA\ at a resolution of 100\,000 with a 
signal-to-noise ratio of 70 but when smoothed to the resolution of the
2010 UVES spectra the UCLES spectrum has a signal-to-noise ratio comparable to
that of the former. 
Further details were given by \citet{rao12}. 

Four spectra were acquired  in February-March 2010
with UVES, the cross-dispersed \'echelle 
spectrograph \citep{dekker00} on the Very Large Telescope of the European Southern Observatory at Cerro Paranal, Chile.
A resolution $R \equiv \lambda/\Delta\lambda \approx 34\,000$ was estimated from
the telluric lines in the  5920\AA\ region. The available spectra cover the wavelength
regions 3300 -- 4500\AA, and 5700 -- 7500\AA. 
In the blue, the  signal-to-noise ratio exceeds 200 at the above wavelengths.  
The spectrum from 2010 February 27 was chosen for an abundance
analysis because lines were least affected by emission in the core and the
absorption line profiles are the most symmetric.
Emission cores  
are prominent in hydrogen Balmer lines, He\,{\sc i} lines
and C\,{\sc ii} lines. 
The emission appears to be variable on a timescale of weeks, 
since the 2010 February 27 spectrum shows little or 
no core emission compared with the 2010 March 2 spectrum.  
Several nebular lines including [O\,{\sc i}],  
[O\,{\sc ii}], [N\,{\sc ii}], and [S\,{\sc ii}] 
are also present \citet{rao13}. 

Spectra at all three epochs are dominated by photospheric absorption. The
vast majority of lines arise from H\,{\sc i}, He\,{\sc i}, C\,{\sc ii},
C\,{\sc iii}, N\,{\sc ii}, N\,{\sc iii}, O\,{\sc ii}, O\,{\sc iii},
Ne\,{\sc i} and Ne\,{\sc ii}. A few lines from other ions are present but
care must be taken to account for blending of these lines with lines
from the above species which dominate the spectrum.  
Lines were identified using the Revised Multiplet Table (RMT) \citep{moore72}, tables of spectra 
of H, C, N, and O \citep{moore93} and the {\it NIST} Atomic Spectra 
Database\footnote{http://www.nist.gov/pml/data/asd.cfm}.

The primary information required from these spectra are the effective temperature and surface gravity 
and the chemical composition
of DY\,Cen at each epoch. 
Quantitative analysis is applied consistently to each spectrum.


\section{Quantitative Fine Analyses}

Determination of the atmospheric parameters and the chemical composition
is based on line-blanketed model atmospheres. The effective temperature
$T_{\rm eff}$ and surface gravity $g$ are obtained from the intersection
of  loci in the $T_{\rm eff}$ versus $\log g$ plane. These
loci include several expressing ionization equilibrium (e.g., C\,{\sc ii}/C\,{\sc iii}
and O\,{\sc ii}/O\,{\sc iii})  and others derived from fits to the
Stark-broadened profiles of He\,{\sc i} lines, as illustrated
below. The microturbulent velocity $\xi$ is determined from O\,{\sc ii}
lines spanning a range in equivalent width. 

In principle, determination of
chemical composition is an iterative process which concludes when the
composition adopted in the computation of the model atmosphere equals the
composition derived from a spectrum.  Test calculations with models computed
for C/He of 0.3 - 1\% and H/He of 0.0001 and 0.1 give essentially the same
atmospheric structure and, thence, the same atmospheric parameters including the
composition.

For DY Cen, photoionization of neutral helium is the principal 
source of continuous opacity in the
optical. Thus, lines of another species, say C\,{\sc ii}, are sensitive to the
C/He ratio. 
The abundances are given as $\log\epsilon$(X)
and normalized with respect to $\log\Sigma\mu_{\rm X}\epsilon$(X) $=$ 12.15
where $\mu_{\rm X}$ is the atomic weight of element X.
With bound-free He transitions providing the opacity, the strength
of the He\,{\sc i} lines is not very temperature sensitive; the
excitation potential of the continuous opacity providing levels
are but slightly elevated with respect to the levels providing the
observed lines. A similar consideration applies to many of the
lines of other ions which have high excitation potentials but
similar to those of He providing the continuous opacity.

Under DY Cen's atmospheric conditions, the approximation of local thermodynamic
equilibrium (LTE) is expected to fail for some species. In recognition of this
failure, the abundance analyses were carried out for non-LTE model atmospheres and for non-LTE
(and LTE) line formation for all major elements and some minor elements. 
Partially-blanketed non-LTE model atmospheres were computed with the 
code TLUSTY \citep{hubeny88,hubeny95} using  atomic data and model atoms  
provided on the
TLUSTY home page\footnote{http://nova.astro.umd.edu/index.html}. 
The microturbulence of $\xi=10\,{\rm km\,s^{-1}}$ was used for computing the
model atmospheres.
These model atmospheres included both bound-free and 
bound-bound transitions of H, He, C, N, O, and Ne in
NLTE. The adopted model atoms,
with their number of levels given in brackets, are:
H\,{\sc i}(9), He\,{\sc i}(14), He\,{\sc ii}(14), C\,{\sc i}(8), C\,{\sc ii}(11), C\,{\sc iii}(12),
C\,{\sc iv}(13), N\,{\sc i}(13), N\,{\sc ii}(6), N\,{\sc iii}(11), N\,{\sc iv}(12), O\,{\sc i}(22),
O\,{\sc ii}(29), O\,{\sc iii}(29), Ne\,{\sc i}(35), Ne\,{\sc ii}(32), and Ne\,{\sc iii}(34).
Other atoms were considered with lines analysed in LTE except for the following for which
TLUSTY provides model atoms with the number of levels in brackets: 
Mg\,{\sc ii}(14), Si\,{\sc ii}(16), Si\,{\sc iii}(12), Si\,{\sc iv}(13), S\,{\sc ii}(14),
S\,{\sc iii}(20), Fe\,{\sc ii}(36), and Fe\,{\sc iii}(50).

Model atmospheres in LTE were computed using TLUSTY and were also taken from
an extensive grid of LTE models described by \citet{behara06}. The latter
models 
were constructed using the fully line-blanketed LTE code {\sc STERNE}
\citep{behara06}, which incorporates Opacity Project bound-free opacities for all important elements
up to and including iron and bound-bound atomic transitions corresponding to some 10$^6$ lines by
means of an opacity-sampling (OS) formalism \citep{behara06}. As a starting approximation, a
composition corresponding to that of \citetalias{jeffery93a} was adopted, with the exception that the iron abundance
was scaled to be solar relative to the silicon abundance. An important feature of the OS approach is
that the microturbulent velocity $\xi$ can easily be adjusted to be consistent with that measured
from the observed spectrum. 
A large value of $\xi$ was found by \citetalias{jeffery93a} and confirmed by the
present analyses 
(see below). A
model grid was computed covering the ranges $T_{\rm eff} = 16\,000 (1\,000) 30\,000$ K 
and $\log g =
1.7 (0.1) 3.0$ cgs. 
Test calculations showed that LTE TLUSTY and LTE STERNE models for
parameters applicable to DY Cen give essentially the same atmospheric
parameters and elemental abundances. It is, thus, very likely that the non-LTE
effects estimated from comparison of results for LTE and non-LTE TLUSTY atmospheres
are representative of those to be found from STERNE models were the latter available
in  a non-LTE variety.

An extensive LTE analysis of the blue optical spectrum of DY\,Cen was reported by \citetalias{jeffery93a}.
Ionization equilibrium was used to establish the effective temperature $T_{\rm eff}$, Stark-broadened
neutral helium line profiles provided the surface gravity $g$, and abundances were obtained for 13 elements 
from H to Fe. 
In addition, the microturbulent velocity $\xi$ and the projected rotation velocity $v_{rot} \sin i$ 
were measured from O{\sc ii} lines.  The approximation of local thermodynamic equilibrium (LTE) for the atomic level populations 
was adopted throughout. 

In the interim it has been shown that departures from LTE strongly 
affect the profiles, and especially the cores, of
neutral helium lines in extreme helium stars of similar 
$T_{\rm eff}$ to DY\,Cen \citep{przybilla05}  and lines of C, N, and O
used to establish atmospheric parameters may also be affected by
departures from LTE. Additionally, 
a complete treatment of line-blanketing is important for the 
temperature stratification and, hence, 
the measurement of $T_{\rm eff}$  of hot He stars \citep{behara06} but
tests show that the line-blanketing adopted for TLUSTY models is
a good approximation to complete line-blanketing.  

In this paper, we have used TLUSTY and SYNSPEC for calculating LTE and
non-LTE model atmospheres and line profiles \citep{hubeny88,hubeny94,hubeny95}.
Stellar atmospheric parameters are
determined from the CASPEC, UCLES and UVES spectra on the assumption
of non-LTE using the H\,{\sc i}, He\,{\sc i}, C\,{\sc ii}, C\,{\sc iii},
C\,{\sc iv}, N\,{\sc ii}, N\,{\sc iii}, O\,{\sc ii} and O\,{\sc iii}
lines. Abundances of elements beyond O are determined from analyses of
lines in the UVES spectra with occasional reference to the CASPEC spectrum and also to the  UCLES
spectrum whose S/N ratio is comparable to that of a UVES spectrum  
when smoothed to the resolving power of a UVES spectrum.
Adoption of the UVES spectrum for the `heavy' element analyses is warranted
 by the superior quality of the UVES spectra relative to the
CASPEC spectrum.
Except where noted, the $gf$-values of the lines are taken from the
NIST database.\footnote{http://www.nist.gov/pml/data/asd.cfm} 
A few other sources consulted for $gf$-values  are referenced in footnotes to the
relevant tables.

For the elements H to Ne, identification of lines suitable for analysis
is not a major issue. In particular, ions of C, N, and O are generally
very well represented and a good selection of clean lines is
available. \citet{moore93} is the primary source of wavelengths and
classifications for these lines. Identifications of the few lines of
`heavy' elements are discussed later. 
Hydrogen Balmer lines are
subject to overlying variable emission. Emission swamps the
absorption component in early Balmer lines: H$\delta$ has a strong emission
core in UVES spectra but strong absorption dominates the line in the
CASPEC and UCLES spectra. Strong variable emission is also seen
in some C\,{\sc ii} lines.
Weak emission
is the only signature of Balmer lines beyond H17 at 3697\AA\
in the UVES spectra. 
Neutral helium lines are well represented. Early lines in several
series exhibit a P Cygni profile, e.g., 5876.6\AA\ and 6678.1\AA.
He\,{\sc i} lines are traceable to the short wavelength limit of the
UVES spectra.  Emission components in H\,{\sc i}, He\,{\sc i},
C\,{\sc ii} and other lines are likely attributable to the star's
extended atmosphere.

\subsection{Non-LTE analyses}

Line profiles and theoretical equivalent widths were obtained for these model atmospheres
using the companion non-LTE code SYNSPEC \citep{hubeny94}. 
Non-LTE abundances were derived  by matching the observed 
absorption profile or its measured equivalent width with the SYNSPEC prediction.
Note that features which are unresolved blends of two or more lines
were synthesized and matched to the observed feature by adjustment of
abundances.

The procedure for determining the $T_{\rm eff}$, $\log g$ and $\xi$ is 
a standard one.
The microturbulent velocity $\xi$  
is estimated from O\,{\sc ii} lines because they show a wide range in equivalent width. 
O\,{\sc ii} lines with similar lower excitation potentials (LEP) were used to minimise the temperature 
dependence:  O\,{\sc ii} lines were used with LEPs about
23, 26, and 28 eV. $\xi$ is found by requiring  the abundance to be 
independent of the measured equivalent width.

For pairs of ions of the same element, insistence upon ionization
equilibrium provides a locus in the ($T_{\rm eff}$, $\log g$) plane.
Available potential
loci include C\,{\sc ii}/C\,{\sc iii}, C\,{\sc ii}/C\,{\sc iv},
C\,{\sc ii}/C\,{\sc iv}, O\,{\sc ii}/O\,{\sc iii}. Not all these loci
are available for all spectra.

An additional locus is provided by fits to the strongest
cleanest He\,{\sc i} line profiles
with their Stark-broadened wings. Predicted line profiles depend on the
electron densities and, therefore, on the temperature and surface gravity.

A final locus is the $T_{\rm eff}$
from photometry. 

The effective temperature and surface gravity
are found  as the best overall fit to the
intersecting loci.

\subsubsection{CASPEC 1987}

From the 1987 CASPEC spectrum, we redetermine the 
stellar parameters ($T_{\rm eff},\log g,\xi$) using the non-LTE model atmospheres 
and the non-LTE line formation code as discussed in the above Section.
As described earlier, $\xi$ is estimated from O\,{\sc ii} lines.
The $T_{\rm eff}$ and $\log g$ are then determined from He\,{\sc i} line profiles, 
the ionization balance for (C\,{\sc ii},C\,{\sc iii}) and the photometric
estimate of $T_{\rm eff}$.
For our analysis, we have used the line list given by \citetalias{jeffery93a} with 
some additions to the C\,{\sc ii} and C\,{\sc iii} lines.
Fits of synthetic spectra convolved with the instrumental profile with a FWHM of 0.2\AA\
according to \citetalias{jeffery93a} give a projected rotational velocity of 20 - 25
km s$^{-1}$ from clean O\,{\sc ii} lines. 

Note that for most of the CNO lines our measured equivalent widths
are in good agreement with those of \citetalias{jeffery93a}. Hence, we have used the \citetalias{jeffery93a} 
equivalent widths and the most recent $gf$-values (see  Section 4).
Our non-LTE analysis gives the final model parameters:
($T_{\rm eff}$, $\log g$, $\xi$)=(19400, 2.1, 20.0), in agreement with
the stellar  LTE parameters derived by \citetalias{jeffery93a}.
The CNO lines given in Table 2 are used in this analysis.
Observed profiles of the He\,{\sc i} 4922\AA, 4471\AA, and 4388\AA\
line are shown in Figure 2 with predicted non-LTE profiles for  non-LTE
atmospheres of $T_{\rm eff}$=19400K and two different surface gravities.
Predicted profiles include the convolution with the instrumental profile and the projected 
rotational velocity.
At this effective temperature, the surface gravity $\log g$ $\simeq$ 2.1 provides an
good fit to these He\,{\sc i} lines. 
For the final model, line by line non-LTE abundances including the mean abundance,
and the line-to-line scatter, are  given in Table 2. The lines giving significantly deviant
abundances are marked by ?, and are not included in estimating the mean.
The abundance rms errors, due to uncertainty in $T_{\rm eff}$ and $\log g$, from
C\,{\sc ii}, C\,{\sc iii}, N\,{\sc ii}, N\,{\sc iii}, and O\,{\sc ii}
are 0.06, 0.24, 0.08, 0.21, and 0.12 dex, respectively.

\subsubsection{AAT/UCLES 2002}

In our analysis of the AAT/UCLES spectrum, we have adopted 
the same procedure and nearly the same lines, as for the CASPEC spectrum; the
spectral bandpasses are almost identical. However, the UCLES spectrum is
generally of higher quality, especially if co-addition of pixels is
employed to reduce the resolving power to that of the UVES spectrum.
The final derived stellar parameters are ($T_{\rm eff}$, $\log g$, $\xi$)=(23000, 2.35, 23.0).
The CNO lines given in Table 2, and the  wings of the observed He\,{\sc i} profiles at
4026\AA, 4922\AA, and 4388\AA\  (Figure 3) are used in this analysis.
Synthetic profiles are convolved with the instrumental profile and with the projected
rotational profile of 20 -- 27 km s$^{-1}$ determined from a fit to clean O\,{\sc ii} lines.
The abundance rms errors, due to uncertainty in $T_{\rm eff}$ and $\log g$, from
C\,{\sc ii}, C\,{\sc iii}, N\,{\sc ii}, N\,{\sc iii}, and O\,{\sc ii}
are 0.05, 0.18, 0.08, 0.18, and 0.01 dex, respectively.

\subsubsection{UVES 2010}

Of the UVES spectra, the spectrum with photospheric absorption lines
least affected by core emissions was used (2010 February 27). 
Table 3 of this paper lists the chosen lines of H, C, N,  and O.

A microturbulent velocity $\xi=24\pm3 {\rm km\,s^{-1}}$ is obtained from
O\,{\sc ii} lines. 
Observed profiles of the He\,{\sc i} 4009\AA\ and 4388\AA\
line are shown in Figure 4 with predicted non-LTE profiles for a non-LTE
atmosphere of $T_{\rm eff}$=25000K and two different surface gravities.
The predicted profiles have been convolved with the instrumental profile 
and the stellar rotation profile. A projected rotational velocity of about
40 km s$^{-1}$
is estimated by using unblended moderately strong lines.
The best-fitting theoretical
profile ($\log g$= 2.50) provides one point on the $T_{\rm eff} - \log g$ locus.
The chosen lines are those least affected by emission.

To the mean of the loci from the He\,{\sc i} profiles, we add several
loci from application of ionization equilibrium to C and O ions and the
$T_{\rm eff}$ from photometry.
Figure 5 shows the several loci. Their intersection suggests that the best non-LTE model
atmosphere has $T_{\rm eff}$=24800$\pm$600K and $\log g$ = 2.50$\pm$0.12. 

Like He\,{\sc i}, H\,{\sc i} line profiles are affected by emissions. 
H\,{\sc i} observed profiles at 3722\AA, 3970\AA\ and 4340\AA\ were chosen for
estimating the NLTE hydrogen abundance by spectrum synthesis.
The line wings of 3970\AA\ and 4340\AA\
profiles are mainly used for this purpose as their cores are severely affected
by emissions. The hydrogen model atoms and
line broadening coefficients are from TLUSTY.
Observed profiles of the 3722\AA, 4397\AA, and 4340\AA\
are shown in Figure 6 with predicted non-LTE profiles for a non-LTE
atmosphere of $T_{\rm eff}$=25000K and $\log g$ = 2.50 for three different 
hydrogen abundances.

The  abundance analysis for all elements was conducted for the model atmosphere
($T_{\rm eff}$, $\log g$, $\xi$)=(25000, 2.50, 24.0). The final photospheric line by line
non-LTE and/or LTE  abundances including the mean abundance,
and the line-to-line scatter, are given in Tables 3 and 4. 
The lines giving significantly deviant
abundances are marked by ?, and are not included in estimating the mean.
The abundance rms errors, due to uncertainty in $T_{\rm eff}$ and $\log g$, from
C\,{\sc ii}, C\,{\sc iii}, C\,{\sc iv}, N\,{\sc ii}, N\,{\sc iii}, O\,{\sc ii},
O\,{\sc iii}, and Ne\,{\sc ii} are 0.11, 0.14, 0.29, 0.09, 0.12, 0.08, 0.14, and
0.13 dex, respectively.

Similarly, we have also conducted the LTE analysis. The LTE TLUSTY models with the LTE line
analysis gives the final model parameters: 
($T_{\rm eff}$, $\log g$, $\xi$)=(24750, 2.65, 30.0). The LTE abundances for the
best LTE TLUSTY model are given in Tables 3 and 4. The abundance rms errors, due to uncertainty 
in $T_{\rm eff}$ and $\log g$, are very similar to those estimated for the appropriate
non-LTE model atmosphere.

\subsubsection{Atmospheric parameters -- Summary}

Results of the analysis of the He, C, N, and O lines for the CASPEC, UCLES and UVES
spectra are summarized in Table 1 and shown graphically
in Figure 5 where the available loci are plotted together with
the $T_{\rm eff}$ from the V-band photometry. The adopted $T_{\rm eff}$ and $\log g$
for each spectrum is shown by the cross in each panel. For a given spectrum,
the cross is a good representation of
each available locus and in  fair to good agreement with the $T_{\rm eff}$ from the
V-band photometry. These spectroscopic  analyses show that DY Cen evolved to higher
effective temperature and higher surface gravities between 1987 and 2010 following the
trend suggested by photometry since early in the twentieth century (Figure 1).

The DY\,Cen's derived stellar parameters for all three epochs (Figure 7) are plotted on
Figure 6 of \citet{saio02} for the merger products. This suggests that DY Cen has evolved at
a constant luminosity corresponding to a 0.9$M_{\odot}$ model starting from 0.6 - 0.5$M_{\odot}$ CO WDs.
DY Cen has become hotter by about 5000 K in 23 years, i.e,
at a rate of about log (dTeff/dt) (K/yr) = 2.34.
This rate is higher when compared with the rates given by \citet{saio02}
in their Figure 7 for the merger products.
\citet{rao12}, based on their radial velocity measurements, noticed that
DY\,Cen is a binary. It is possible that the binary companion, by interaction, is increasing
DY\,Cen's mass loss rate, and hence, enhancing its evolutionary rate as observed \citep{schoenberner79}.

\section{Chemical Composition}

Since the dominant opacity across the optical is provided by neutral
He atoms, lines of an ion are sensitive to the abundance ratio of that element to
helium. Abundances for element X -- $\log\epsilon$(X) --  are normalized to  
$log\Sigma\mu_{\rm X}\epsilon$(X) $=$ 12.15 where $\mu_{\rm X}$ is the atomic 
weight of element X.

With the exception of neon which is well represented by both Ne\,{\sc i}
and Ne\,{\sc ii} lines, the elements beyond C, N, and O provide 
either no 
or just a few detectable lines. Thus, the accuracy of an elemental abundance 
depends in large part on secure identifications of stellar lines.
In this regard, the higher quality of the UVES and (smoothed) UCLES
spectra relative to the CASPEC spectrum has led to several revisions
of the line identifications used by \citetalias{jeffery93a} in
their abundance analysis.

In the following, selected elements from H to Fe are
discussed and recommended  abundances derived. Recommendations are given in Table 5 where we also
list the abundances from \citetalias{jeffery93a} (their Table 2). 
The various RMT cited refer to \citet{moore72}

{\bf Hydrogen:} The H abundance from select Balmer lines in the
UVES spectra is given in Table 3: the mean non-LTE abundance is 10.65
corresponding to a H/He ratio of 0.13 by number.  
Our analysis of H$\beta$, H$\gamma$ and H$\delta$ in the
CASPEC and UCLES spectra confirm this abundance with 10.6$\pm$0.1 and 10.7$\pm0.1$ from
CASPEC and UCLES spectra, respectively. 
These H abundances are consistent with the LTE value of 10.76$\pm$0.20
obtained by \citetalias{jeffery93a}; the corrections for non-LTE effects are small.
With this H/He ratio, DY Cen is among the most H-rich of the He-rich H-poor supergiants
including the EHe and RCB stars, as \citetalias{jeffery93a} recognized.

{\bf Carbon:} Carbon is represented by roughly equal numbers of C\,{\sc ii} and C\,{\sc iii}
lines and in the case of the UVES spectra by two C\,{\sc iv} lines. Examination of the non-LTE
abundances from the different ions in Tables 2 and 3 shows consistent results from the
ions in the individual spectra. The unweighted means that suggest a quasi-constant abundance across the
CASPEC, UCLES and UVES spectra: 9.54 (CASPEC), 9.41 (UCLES) and 9.74 (UVES) for mean abundances of
9.57 from C\,{\sc ii} and 9.54 from C\,{\sc iii} lines. Given the errors of measurement and the
line variability including appearance of emission components, the apparent variation of the C
abundance is not considered a real effect. We adopt a mean C abundance of 9.55 or a C/He ratio of
1.0\% by number. 

{\bf Nitrogen:} The N abundance is heavily dependent on the N\,{\sc ii} lines with a single
N\,{\sc iii} line providing supporting evidence. The two ions give very similar N
non-LTE abundances, a comforting result in that the non-LTE corrections are quite
different for the two ions. The mean abundances do not change significantly from the
CASPEC, UCLES and UVES spectra: $\log\epsilon$(N)= 7.78 is adopted. 

{\bf Oxygen:} For the UVES spectra, the O abundance is based on roughly equal numbers of
O\,{\sc ii} and O\,{\sc iii} lines with good agreement between the non-LTE abundance from the two ions.
The CASPEC and UCLES spectra from O\,{\sc ii} lines alone give a slightly lower O abundance than 
the value from the UVES spectrum.  The mean non-LTE abundance from the three sets of O\,{\sc ii} lines 
$\log\epsilon$(O)= 8.87.

{\bf Fluorine:} A multiplet by multiplet search for F\,{\sc ii}
lines resulted in detections of two lines from RMT 3 (Figure 8): the strongest
line, a blend of three components, is present and appears unblended
at 3505.6 \AA; the next strongest line, also a blend of three
components, at 3503.1 \AA\ is present; the multiplet's weakest
line at 3501.4 \AA\ falls in the wing of a He\,{\sc i} line. Lines
from other multiplets are masked 
by stronger lines from other contributing ions. 
Table 4 shows that the two F\,{\sc ii} blends give similar LTE
abundances with $gf$-values from the NIST website. The abundance 
$\log\epsilon$(F) = 7.0 is adopted.
(F\,{\sc i} lines are absent as expected - see \citet{pandey06c}'s collection of spectra of
extreme Helium stars.) 

{\bf Neon:} Neon is well represented by both Ne\,{\sc i} and
Ne\,{\sc ii} lines. Model atoms in SYNSPEC allow us to compute
non-LTE (and LTE) Ne abundances for both the neutral and ionized
lines.

Neutral Ne is represented in DY Cen by several multiplets in the red. 
Five  clean
lines are given in Table 4 with $gf$-values from the NIST website. 
The equivalent widths of these lines are
substantially smaller than those reported by \citetalias{jeffery93a} and
the stellar profiles appear to be composed of a blend of two equal
components or contaminated by weak central emission.
The mean Ne abundance from Ne\,{\sc i}  lines is 8.7 and 9.5 from non-LTE
and LTE analyses, respectively. 

Ionized Ne provides many lines and a large selection of the cleanest
lines is provided in Table  4. 
The $gf$-values are taken from NIST website if available or else from Kurucz's website.
For the UVES spectrum, the mean
neon abundance from {\bf Ne\,{\sc ii}} lines is 8.4$\pm$0.2 (LTE) and
8.0$\pm$0.1 (non-LTE). These mean values are substantially lower than
from the Ne\,{\sc i} lines: the difference is 0.7 dex for the
non-LTE analyses. The Ne abundance from the Ne\,{\sc i} lines is
confirmed by reanalysis of the Ne\,{\sc i} lines measured by \citet{rao1993}
and listed by \citetalias{jeffery93a}
off a spectrum from 1989.

{\bf Magnesium:} In the bandpass of
the UVES spectrum, the sole signature of magnesium
is the Mg\,{\sc ii} 4481\AA\ feature. 
Atomic data for the feature composed of three unresolved  lines are
from \citet{kelleher2008}'s
critical evaluation which also appears on the NIST website. The Mg
abundance from the UVES spectrum is $\log\epsilon$(Mg) = 6.8 and
7.1 for non-LTE and LTE analyses, respectively.  

The 4481\AA\ feature is also present on the CASPEC and UCLES spectra
where its equivalent width is 256 m\AA\ and 176 m\AA, respectively,
to be compared with 154 m\AA\ from the UVES spectrum. The feature
when analysed with the appropriate model atmosphere (Table 1) gives very
similar non-LTE Mg abundances: namely, 6.70, 6.67 and 6.76 from the
CASPEC, UCLES and UVES spectra, respectively.

{\bf Aluminium:} The aluminium
abundance must be determined from the few available 
Al\,{\sc iii} lines  with
$\log gf$-values
taken from \citet{kelleher2008al}'s compilation except that values for
RMT 8 at 4480\AA\  not included by them are taken from Kurucz.  
Al\,{\sc ii} and A\,{\sc iv} lines are not expected to
be detectable in the available spectra, an expectation confirmed
by examination of the spectra.  

For the UVES spectra, the Al abundance comes from the four features in
Table 4.
The Al LTE abundances from three detected lines 
are 6.07 from 3601.6\AA, 7.70 from 5722.7\AA\  and  6.22 from the 
blend at 4480\AA.  The large 
discrepancy between the 3601\AA\ and 5722\AA\ lines comes from
two well resolved lines and no known blend for either line. 
This discrepancy suggests serious non-LTE
effects are present. A feature at 4149.9 \AA\ is absent and sets an
abundance upper limit of 5.1, a limit more than 1 dex below the
abundance provided by the 3601.6 and 4480 \AA\ lines. The 4149.9 \AA\
feature is also not detectable on the CASPEC and UCLES spectra.
 Clearly, this limited
set of Al\,{\sc iii} lines gives apparently inconsistent Al abundances.
Our adopted abundance is 6.1 based on the consistent 3601.6 and 4480 \AA\
features.  

\citetalias{jeffery93a} chose three Al\,{\sc iii} features
from two multiplets with a
third multiplet rejected because of a blend with an O\,{\sc ii}
line. In addition to the 4480 \AA\ feature, they adopted RMT 3 with
a line at 4512.6 \AA\ and a blend  near 4529 \AA. Remeasurement and
reanalysis  of these
three features from CASPEC and UCLES spectra confirms the 1993 result
that they give consistent Al abundances and suggests that our adopted
abundance is a reasonable choice.

{\bf Silicon:} At the atmospheric conditions prevailing from
1987 to 2010, one expects to see lines from the ions Si$^+$, Si$^{2+}$
and Si$^{3+}$ depending on wavelength coverage of available
spectra. Atomic data including $gf$-values are taken from the
critical evaluation by \citet{kelleher2008si}. 

Si II: The only Si\,{\sc ii} lines on the CASPEC spectrum are from RMT 3 at
4130.89\AA\ and 4128.05\AA\ with $\log gf$-values of 0.57 and 0.36,
respectively. \citetalias{jeffery93a} reported these as
absorption lines with equivalent widths of 273m\AA\ and 268m\AA,
respectively. Inspection of these lines on the UVES spectra shows that
both are blended to the red with a stronger O\,{\sc ii} line; the Si\,{\sc ii}
equivalent widths attributed to the CASPEC spectrum probably refer to the
blend of Si\,{\sc ii} and O\,{\sc ii} lines. Synthesis of the blends
in the UVES spectrum gives the non-LTE Si abundance of 8.4 from both lines.  

An advantage of the UVES spectra over the CASPEC and UCLES spectra is that 
greater wavelength coverage to the red includes 
Si\,{\sc ii} lines from RMT 2 (6347\AA\ and 6371\AA),
RMT 4 (5978\AA\ and 5957\AA). These lines are in emission and not in
absorption; the upper term of RMT 2 is the lower term for RMT 4. 
Application of LTE or non-LTE to absorption lines at 4128\AA\ and
4130\AA\
is of suspect validity given that these red multiplets appear in emission.

Si III: A search of the UVES spectra  provided the well-resolved
line at 5739.7\AA\ (RMT 4) with
no obvious blends -- see Table 4 for the LTE and non-LTE Si
abundances. Another line at 3791.4\AA\ is blended with an O\,{\sc ii}
line. Assuming that this line is solely due to Si\,{\sc iii} gives
Si abundances of 7.03 and 7.14 for LTE and non-LTE
analyses, respectively. These values are about 0.6 dex less than
the abundances from the 5739\AA\ line. Inclusion of the O\,{\sc ii}
contribution to the 3791\AA\ line obviously increases the abundance
difference between the 5739\AA\ and 3791\AA\ lines.  

Two additional multiplets are present on the CASPEC and UCLES
spectra and provide six clean lines well suited to abundance
analysis: three lines of RMT 2 at 4552.6, 4567.8 and 4574.7\AA\ and
three lines or blends of RMT 9 at 4829.1, 4819.7 and 4813.3\AA. 
(\citetalias{jeffery93a} did not include these lines in their
analysis.)  These features give consistent abundances when analysed with the
appropriate model atmosphere: the mean
non-LTE Si abundances are 6.86 (RMT 2) and 6.84 (RMT 9) for the
CASPEC lines  and 6.92 (RMT 2) and 6.75 (RMT 9) for the UCLES lines for
the grand mean of 6.84.

Si IV:  The two Si\,{\sc iv} lines of RMT 1
were detected and analysed by \citetalias{jeffery93a}. The weaker line 
at 4116.104\AA\ appears unblended. The stronger line at 4088.86\AA\ is
blended with an O\,{\sc ii} line from RMT 48, as noted by Jeffery \&
Heber. These lines which appear also in the
UCLES and UVES spectra are the only
representatives of the Si$^{3+}$ ion in the spectra. 

Analysis of the 4116\AA\ line on the UVES spectrum  gives 
Si abundances of $\log\epsilon$(Si) = 7.03 (LTE) and 6.69 (non-LTE).

A consideration of the Si\,{\sc ii}, Si\,{\sc iii} and Si\,{\sc iv}
lines discussed in this section suggests that the Si abundance is 
$\log\epsilon$(Si) $\simeq$ 6.8 according to the suite of CASPEC,
UCLES and UVES spectra. In drawing this conclusion, we give
zero weight to the Si abundance from the Si\,{\sc ii} 4128\AA\ and
4130\AA\ on the grounds that Si\,{\sc ii} lines in the red are in
emission which feature is not accounted for by our non-LTE
analysis. A concern is that the clean Si\,{\sc iii} 5739\AA\ line
gives a higher abundance of 7.2 and the blended Si\,{\sc iii} 3791\AA\ line 
gives the upper limit of 6.6 when the line is assumed to be purely from
the Si$^{2+}$ ion.

{\bf Phosphorus:} A search of UVES spectra was conducted 
for lines of P\,{\sc ii},
P\,{\sc iii} and P\,{\sc iv} lines. No lines of P\,{\sc ii} were
found.  Two lines of P\,{\sc iii} and three lines of P\,{\sc iv}
were deemed potential suitable for abundance analysis.
Three lines of P\,{\sc iii}
were the basis for the P abundance
reported by \citetalias{jeffery93a}. 

P III: Two lines 
appear to be unblended: 4080.09\AA\ from RMT 1 and
4222.20\AA\ from RMT 3. A second line from RMT 3 at 4246.72\AA\ line 
was listed by \citetalias{jeffery93a} but this is blended with 
C\,{\sc iii} and N\,{\sc ii} lines.
NIST website's $\log gf$-values are from
the evaluation by Wiese et al. Recent quantum calculations by
\citet{froese2006} give slightly different results:
the 2006 values are 0.17 dex and 0.08 dex smaller for RMT 1 and 3, respectively.
We adopt the 2006 calculations. 
Abundances are LTE estimates because TLUSTY does not include model P
ions.

 In the UVES spectrum, the 4080\AA\ line has a much smaller
equivalent width than the 4222\AA\ lines (20m\AA\ versus 157m\AA) and yields
a lower P abundance (5.50 versus 5.99). The UCLES spectrum confirms this
equivalent width difference. 
Yet, the equivalent widths of the 4080\AA\ and 4222\AA\ lines
listed by Jeffery \& Heber from their CASPEC spectrum are 
almost identical. 
Examination of the CASPEC spectrum suggests that
the 4080\AA\  feature measured previously was displaced by about 0.5\AA\ from
the expected wavelength.
In light of the agreement between UVES and UCLES spectra, it would appear that
the LTE analysis results in approximately a 0.5 dex abundance difference
between the two  lines. It does not appear that the 4222\AA\ line is 
contaminated by an unidentified line or that the 4080\AA\ is weakened
by emission. Perhaps, a non-LTE analysis would eliminate the 0.5 dex
difference.

P IV: Lines of RMT 1 and 2 appear to be present as weak lines. 
the strongest two lines of RMT 1 --  3347.72\AA\ and 3364.44\AA\ -- are
present as is the single line of RMT 2 at 4249.57\AA. The latter line is
too weak to be detected on either the CASPEC or UCLES spectra. The
former two lines fall outside the CASPEC and UCLES bandpasses.
 The $\log gf$-values
given by NIST are quantum calculations \citep{zare1967}. More recent calculations
\citep{froese2006} giving similar results are adopted in Table 4.

With the exception of the low P abundance from the P\,{\sc iii} 4080\AA\
line, a P LTE abundance of 6.0 is indicated from one P\,{\sc iii} and
three P\,{\sc iv} lines.

{\bf Sulphur:} The S abundance
given by \citetalias{jeffery93a} was based on three S\,{\sc ii}
and two S\,{\sc iii} lines. A reassessment of sulphur's contribution to
DY Cen's spectrum was made using the UVES spectra. Atomic data were taken from
\citet{Podobedova09}'s critical evaluation.

S II: Inspection of the UVES spectra gave no convincing identifications of
S\,{\sc ii} lines. The likely strongest line  -- 5453.855\AA\ from RMT 6 --
falls outside the bandpass of all available spectra. Several lines are not
detectable on the UVES and/or the UCLES spectra and provide upper limits to the
S abundance consistent with the abundance provided by the S\,{\sc iii} lines.
Three S\,{\sc ii} lines were measured by \citetalias{jeffery93a}: two lines --
4815.6\AA\ and 4716.3\AA\ from RMT 9 -- and one -- 4162.7\AA\ -- from RMT 44.
On the UVES and UCLES spectra, the 4815.6\AA\ line is absent, the 4716.3\AA\ may
be present and the 4162.7\AA\ line is blended with a C\,{\sc iii} line.

S III: The ion S$^{2+}$ is represented by a handful of weak lines. The two
moderately strong lines from RMT 4 listed by Jeffery \& Heber are badly blended.
Table  4 details the selected lines
providing abundance estimates; other lines from some of the multiplets are
obviously present and might provide an abundance estimate were synthetic
spectra computed. No  line in Table 4 is of a strength to be detectable
in the CASPEC spectrum.  

The mean non-LTE sulphur abundance from the five S\,{\sc iii} lines is
$\log\epsilon$(S) = 6.2 after a negligible correction for non-LTE
effects. 
 
{\bf Argon:} Leading lines of the lowest multiplets of Ar\,{\sc ii}
are absent. In 
particular, 4400.986\AA\ from RMT 1 with $\log gf = -0.28$ is not detectable
on the UVES spectra  and 4806.020\AA\
with $\log gf = 0.21$ is not present on the UCLES spectrum.
The LTE Ar abundances are $\log\epsilon$(Ar) $\leq$ 7.2 and 6.6
from 4400\AA\ and 4806\AA, respectively. Our equivalent width limit for
4806\AA\  of 15 m\AA\ is similar to the 
measurement of 10m\AA\ reported by \citetalias{jeffery93a}
from their CASPEC spectrum.  All Ar\,{\sc i} lines within
the bandpasses of available spectra are far below detection limits.

{\bf Iron:} \citetalias{jeffery93a} used two
Fe\,{\sc iii} lines from RMT 4 to derive the low Fe abundance
$\log\epsilon$(Fe) = 5.04, i.e., 
2.4 dex below the solar abundance. 
Our search of the UVES spectrum yielded upper limits and one weak line 
from multiplets 4, 36, 45 and 118.
The $\log gf$-values are taken from Kurucz.

The leading line of RMT 4 at 4419.596 \AA\ is coincident with an
emission line with emission of comparable strength present on the
UCLES spectrum and just possibly on the CASPEC spectrum. Emission 
may be present also at 4431.019 \AA\ at a strength too weak to
be seen in the UCLES spectrum: an absorption equivalent width is no
stronger than 13 m\AA\ or the LTE abundance is less than $\log\epsilon$(Fe)
= 7.0. This is one of the two lines used by Jeffery \& Heber who gave the
equivalent width as 67 m\AA. Their other line at 4395.755 \AA\ is blended
with an O\,{\sc ii} line.  The RMT 4 line at 4352.577 \AA\ is absent
with an equivalent width limit of 6 m\AA\ yielding the non-LTE Fe
abundance limit $\log\epsilon$(Fe) $\leq$ 7.2. 

Multiplet RMT 36  provides a possible detection of the multiplet's second
strongest line: the line at 3603.888 \AA\ has an equivalent width of
13 m\AA\ and gives the non-LTE abundance of $\log\epsilon$(Fe) = 6.0.
Multiplets 45 and 118 provide useful upper limits to the Fe
non-LTE abundance (see Table  4) in the range 6.0 to 6.4.
In short, the adopted Fe non-LTE abundance is $\log\epsilon$(Fe) $\simeq$
6.0. 

A nagging concern about the abundance analysis is the 0.7 dex difference
between the non-LTE Ne abundance provided by the red Ne\,{\sc i} and blue-ultraviolet
Ne\,{\sc ii} lines. The UVES spectra are the only available spectra of DY Cen to provide
both Ne\,{\sc i} and Ne\,{\sc ii} lines. The Ne\,{\sc i} lines analysed by JH93 confirm
our Ne abundance from these lines. One might attribute the 0.7 dex difference to an inadequate
treatment of non-LTE effects in line formation in an atmosphere that fails to resemble the chosen 
theoretical atmosphere. One cannot help but notice that the red-Ne\,{\sc i} - blue-Ne\,{\sc ii}
abundance difference is reproduced apparently in abundances from the selection of Al\,{\sc iii} lines 
and of Si\,{\sc iii} lines (see Table 4). Are these differences unrelated non-LTE effects? Or is
there a common wavelength effect such as an error in the modelling of the continuous opacity?

\section{Commentary on DY Cen's composition}

Table 5 summarizes the abundances. Columns two and three give our
non-LTE and LTE abundances, respectively and previous estimates by
\citetalias{jeffery93a} are given in column four. Composition of the
solar photosphere is given in the final column \citep{asplund09}.

{\bf Comparison with Jeffery \& Heber:}

A fair comparison involves the LTE abundance estimates.
Inspection of Table 5 shows that the two determinations are in good agreement
(i.e., differences of less than $\pm$0.2 dex with the difference for
C at 0.3 dex) except for three elements: -1.1 (Si), -0.9 (S) and +1.6 (Fe)
where the numerical value is our LTE abundance minus their LTE abundance
in dex. The principal reason for these differences appear to
be found in differing choices of lines (see above). 

{\bf Initial metallicity:} DY Cen's initial metallicity should be judged
from elements most likely unaffected by nucleosynthetic processes
whose products are now in the star's atmosphere. This criterion identifies
the sequence of heavy elements from Mg to Fe. For elements for which a non-LTE
abundance was determinable, the differences (in dex) between abundance for
DY Cen and the solar photosphere \citep{asplund09} are
-0.9 (Mg), -0.7 (Si), -0.9 (S), and -1.5 (Fe). A straight average of these
four would give a metal deficiency of -1.0 dex.   The lower Fe abundance
could point to the origin of DY Cen in the Galactic thick disk where
Mg, Si, and S are overabundant relative to Fe by about 0.3 dex and, hence,
an iron abundance of about -1.2 dex.

For the remaining  heavy elements with a LTE abundance estimate only,
the differences with the solar photosphere are somewhat confusing for
Al and P:
-0.3 (Al) and +0.6 (P) but, as discussed above, the available lines for
both elements do not provide an entirely consistent set of LTE
abundances. Non-LTE corrections for Al seem likely to drive the Al
abundance downward and provide a difference with the solar photosphere
more in line with the values for Mg, Si and S. For an iron
deficiency of 1.0 dex, the expected P abundance for the thick disk is
about 4.7 \citep{caffau11}, i.e., a non-LTE correction of -1.3 dex
is implied or the chosen P\,{\sc iii} and P\,{\sc iv} lines are
seriously blended.

{\bf Fluorine:} The fluorine abundance is in line with the abundance 
found for extreme helium stars (EHes) by \citet{pandey06c} from F\,{\sc i}
lines and for many RCB stars also from F\,{\sc i} lines by Pandey,
Lambert \& Rao (2008). The overabundance relative to the solar photosphere
is 2.5 dex or 2.7 dex if the recent redetermination of the solar F
abundance is adopted \citep{maiorca14}. 

In the case of the EHes and RCrBs, synthesis of F is identified with
a hot phase as two low mass white dwarfs merge -- the double-degenerate
(DD) scenario. In the competing scenario for forming a EHe or a RCrB,
a  low mass AGB experiences a final or a late He shell flash but F
synthesis is not expected in this case. This latter expectation seems
confirmed by the case of Sakurai's object (V4334 Sgr) for which
\citet{pandey08} found an upper limit for the F abundance of 5.4 dex
or about 1.6 dex less than the typical F abundance of EHes and most
RCrBs. 

{\bf Comparison with EHes:} Spectral classification supported by the general
characteristics of the chemical composition suggest that DY Cen's origin is
closely related to that of the Extreme Helium stars (EHes) whose origins
in turn are supposed to be related  to those of the R Coronae Borealis
(RCB) stars. Compilations of the compositions of EHes and RCBs 
\citep{pandey06a,pandey11,jeffery11a}
show that key nucleosynthetic
signatures of EHes are shared with DY Cen, i.e., DY Cen has the typical C/He ratio ($\sim 1$\%)
and marked Ne and severe F overabundances  of EHes. Similarity of nucleosynthetic
signatures encourages the view that DY Cen and EHes share a common origin -- 
a merger of a He and a C-O white dwarf (i.e., the Double-Degenerate (DD) scenario). 
This view is, perhaps, challenged by the
exceptional H abundance of DY Cen which is 2 dex greater than the next most H-rich
EHe and less seriously tested by DY Cen's Fe abundance which is among the lowest found for
EHes. A severe challenge to the idea that DY Cen is a product of the DD scenario is the 
proposal by \citet{rao12} that the star is a spectroscopic binary with a low mass
unseen secondary which might be a He white dwarf or a low-mass (stripped?) main sequence
star. Of course, one may suppose that DY Cen was originally a triple system which, thanks to a
merger of two stars, is now a binary star.   

\citet{rao12} suggest that DY Cen is a common-envelope system with the secondary
presently embedded within the primary's envelope. Theoretical studies of the formation of 
H-poor stars in binaries
(e.g., \citet{pod08}) sketch how binary systems of two normal stars may evolve
through a common-envelope to either single or binary H-poor compact star and evolving to a sdB star,
a hot compact H-poor star approaching a white dwarf cooling track. Atomic diffusion alters the
surface composition of a sdB star and, therefore, it is difficult to correlate sdB compositions
with the composition of putative predecessors.  Common-envelope systems leading to a
sdB in a binary (e.g., \citet{pod08}, Figure 2),
 in contrast to merger scenarios, do not
experience nucleosynthesis resulting in overabundances of Ne and, in particular, F. Thus, it
would appear that a common envelope is insufficient to account for DY Cen and a merger or similar
event accompanied by nucleosynthesis is a necessary part of DY Cen's history. Nonetheless, it should be noted
that the majority of sdB stars are binaries, and several have periods in the 10-100 day range
\citep{maxted01,copper11,barlow12}.

\section{Concluding Remarks}

Judged solely by effective temperature, surface gravity and chemical composition, DY Cen is
an Extreme Helium star, albeit one with an unsually large amount of hydrogen. This identification of
DY Cen as an EHe is in conflict with the identification of the star as a spectroscopic binary \citep{rao12}
and both the generally accepted idea that EHes form by the merger of two white dwarfs and
the observational (if tentative) conclusion from radial velocity studies that (other) EHes are single
stars. Resolution of the conflict is expected to come from an intensive spectroscopic campaign
covering about 120 days -- thrice the length of the orbital period found by Rao et al. Velocity
and profile variations over this campaign should tease out the orbital velocity variation from those
arising from atmospheric pulsations and wind instabilities.

\acknowledgments

We thank the referee Geoff Clayton for the nice report.
We would like to thank Anibal Garc{\'{\i}}a-Hern{\'a}ndez
for his support. The UVES observations were obtained with
his collaboration. GP and NKR would like to thank Simon Jeffery
for his hospitality at Armagh Observatory during 2013 January to
February when a part of this work was done. 
We thank Vincent Woolf, who obtained and reduced the 2002 AAT/UCLES spectrum.
DLL acknowledges the support of the Robert A. Welch Foundation of Houston, Texas through grant F-634.



\begin{figure}
\epsscale{1.00}
\plotone{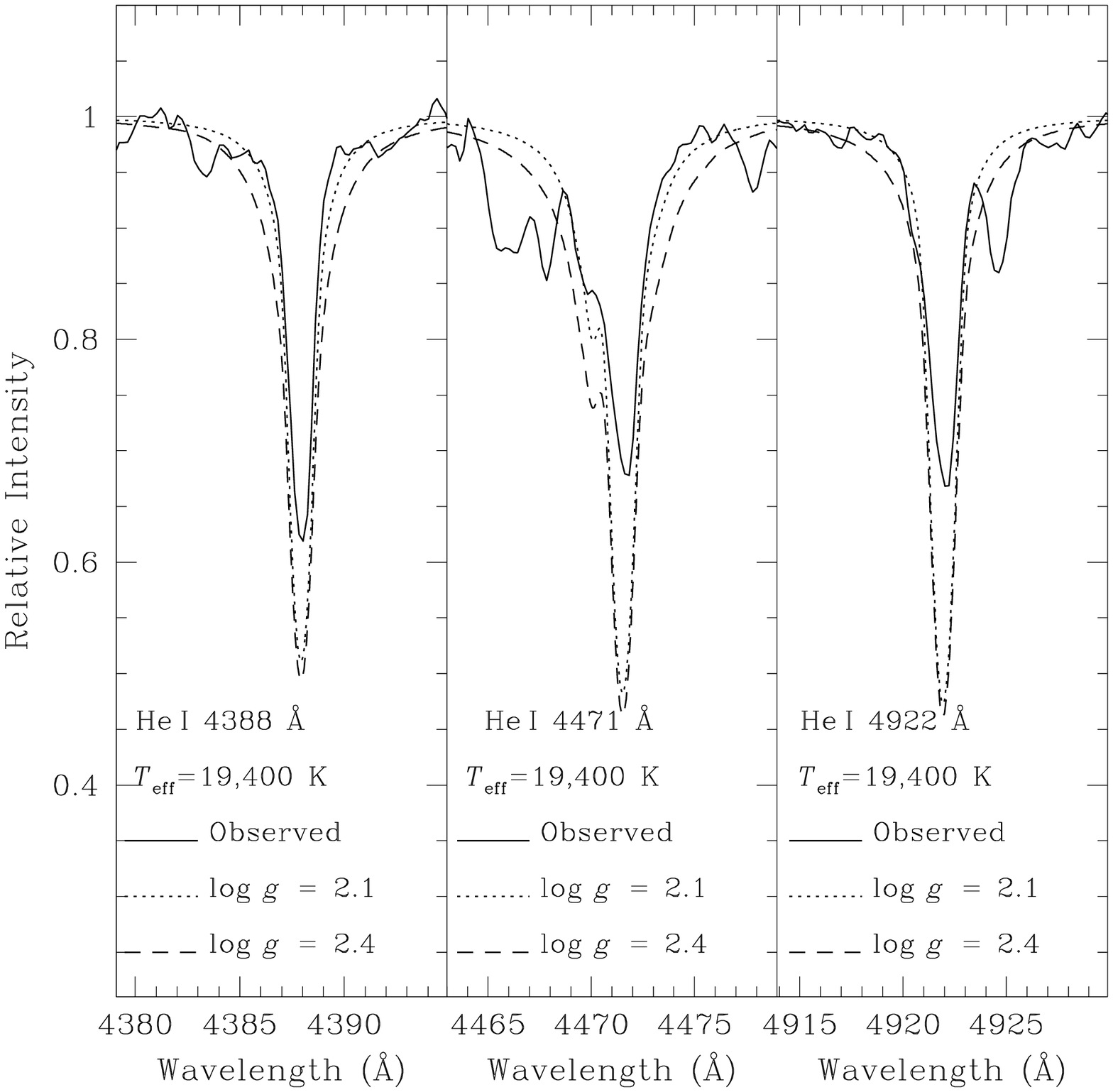}
\caption{The observed CASPEC spectrum and theoretical NLTE He\,{\sc i} line
profiles calculated using the
NLTE model $T_{\rm eff}$=19400 K, for two different $\log g$ values $-$ see key
on the figure. \label{f:he1.nlte}}
\end{figure}

\begin{figure}
\epsscale{1.00}
\plotone{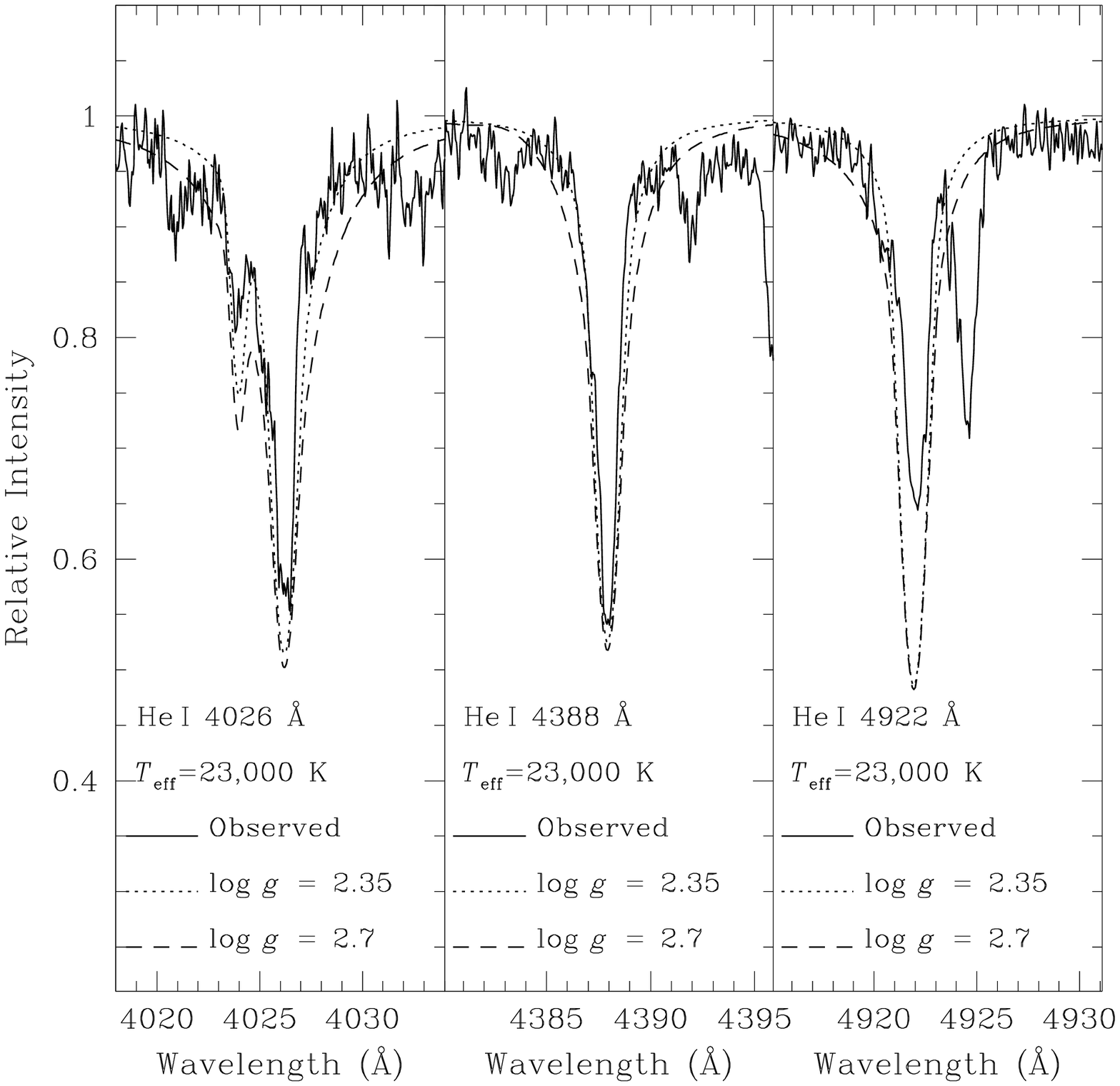}
\caption{The observed UCLES spectrum and theoretical NLTE He\,{\sc i} line
profiles calculated using the
NLTE model $T_{\rm eff}$=23,000 K, for two different $\log g$ values $-$ see key
on the figure. \label{f:he1.nlte}}
\end{figure}

\begin{figure}
\epsscale{1.00}
\plotone{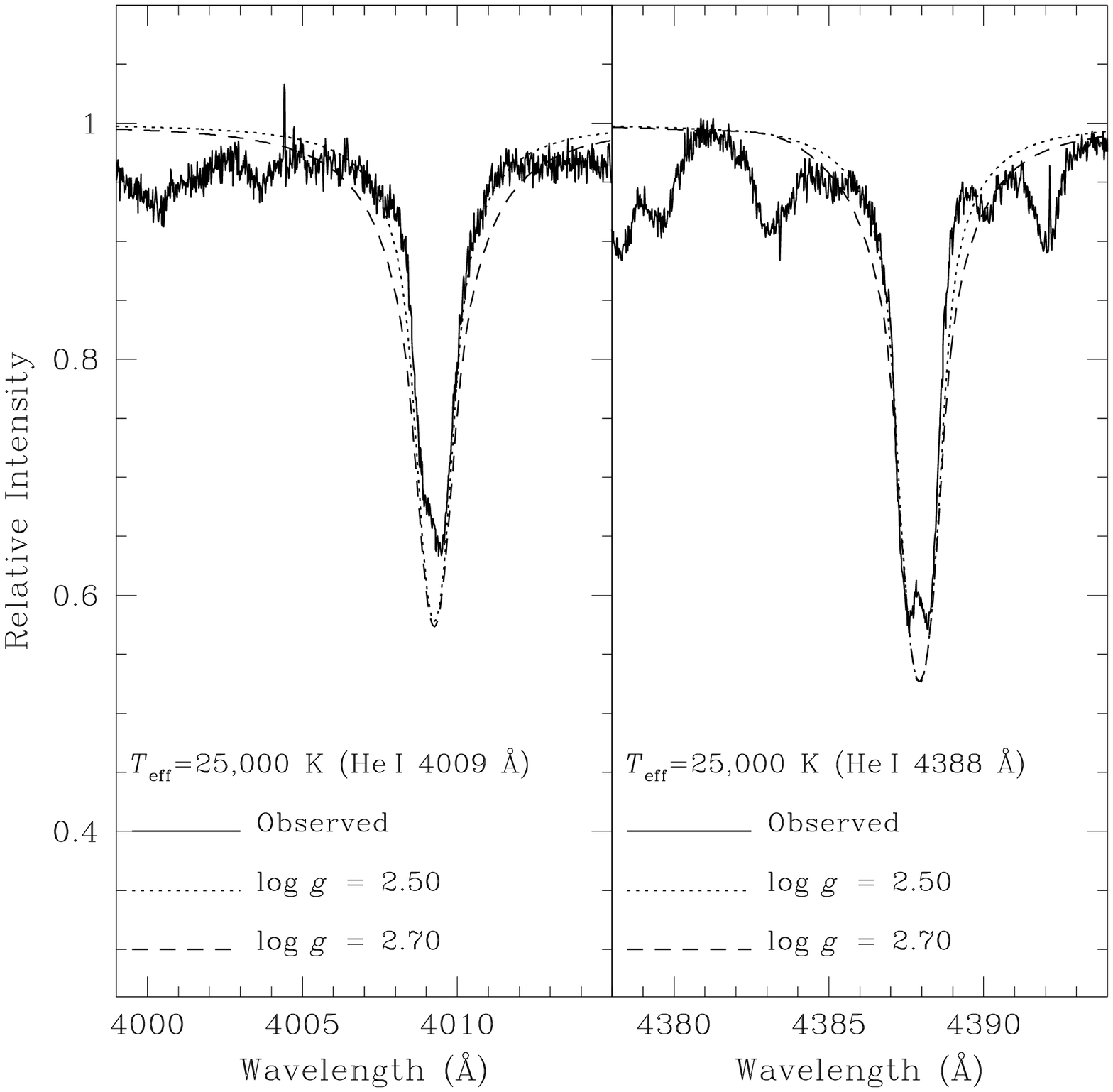}
\caption{The observed UVES spectrum and theoretical NLTE He\,{\sc i} line
profiles calculated using the
NLTE model $T_{\rm eff}$=25,000 K, for two different $\log g$ values $-$ see key
on the figure. \label{f:he1.nlte}}
\end{figure}




\begin{figure}
\epsscale{1.00}
\plotone{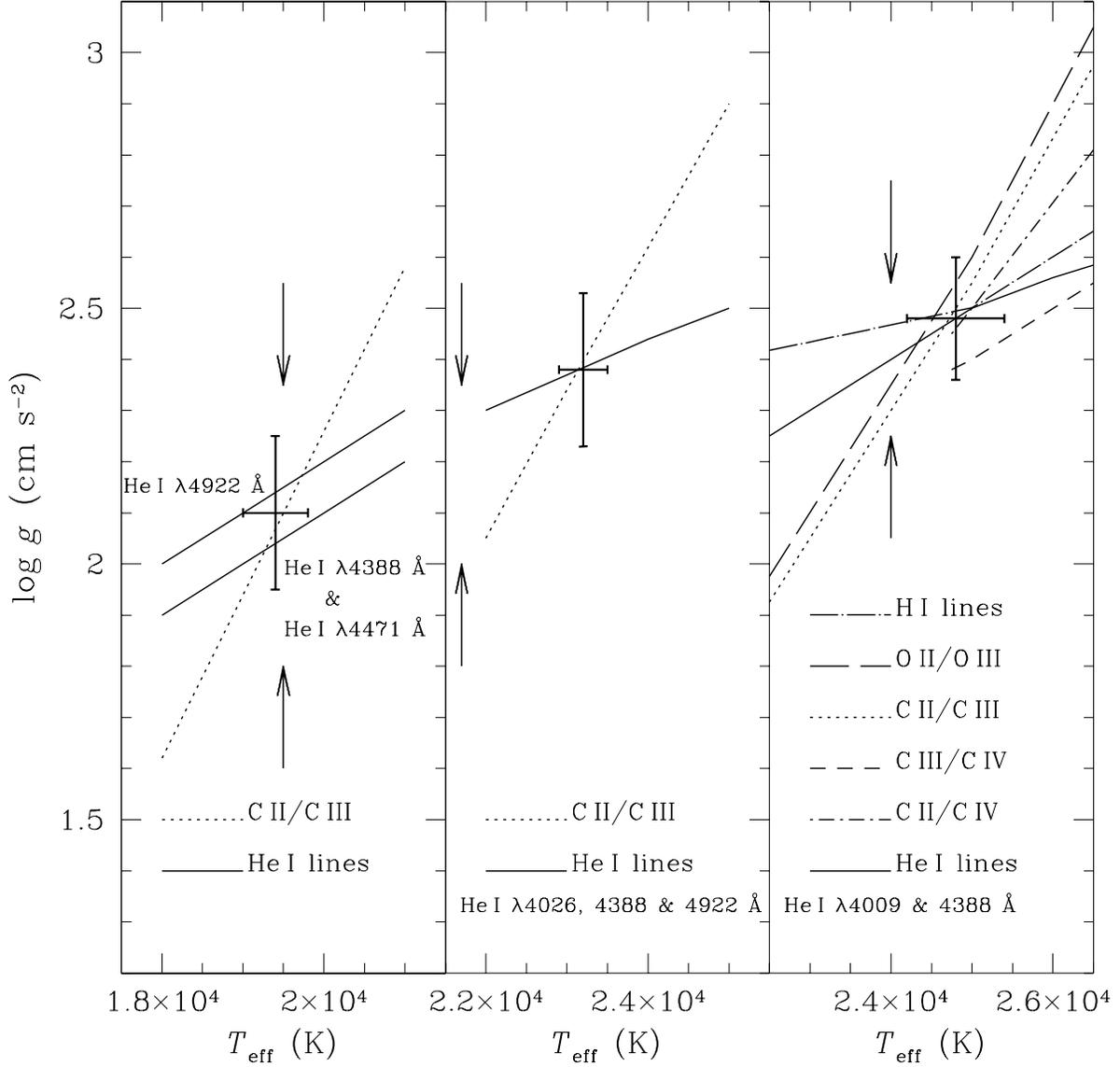}
\caption{The $T_{\rm eff}$ vs $\log g$ plane for 
CASPEC, UCLES, and UVES spectra $-$ from left to right.
Loci satisfying ionization equilibria are plotted $-$ see keys on the figure.
The loci satisfying  optical He\,{\sc i} line profiles
are shown. The locus satisfying optical H\,{\sc i} line profile
is also shown for the UVES spectrum. The cross shows the adopted 
NLTE model atmosphere parameters. The $T_{\rm eff}$ derived from photometry
is shown by arrows. \label{f:gt.nlte}}
\end{figure}

\begin{figure}
\epsscale{1.00}
\plotone{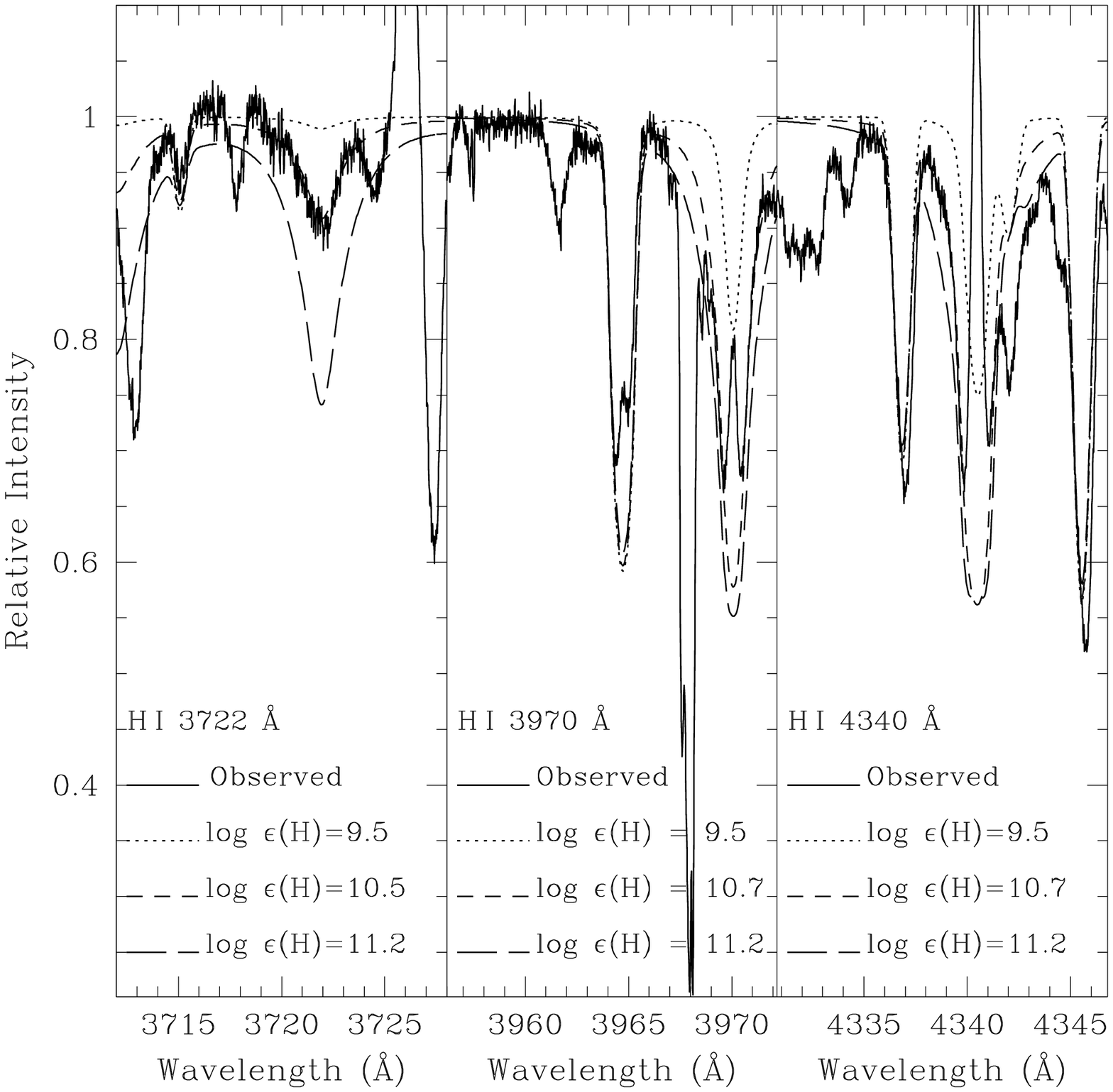}
\caption{The observed UVES spectrum and theoretical NLTE H\,{\sc i} line
profiles calculated using the
NLTE model $T_{\rm eff}$=25,000 K and $\log g$ = 2.5, for three different
H abundances $-$ see key on the figure. \label{f:h.nlte}}
\end{figure}

\begin{figure}
\epsscale{1.00}
\plotone{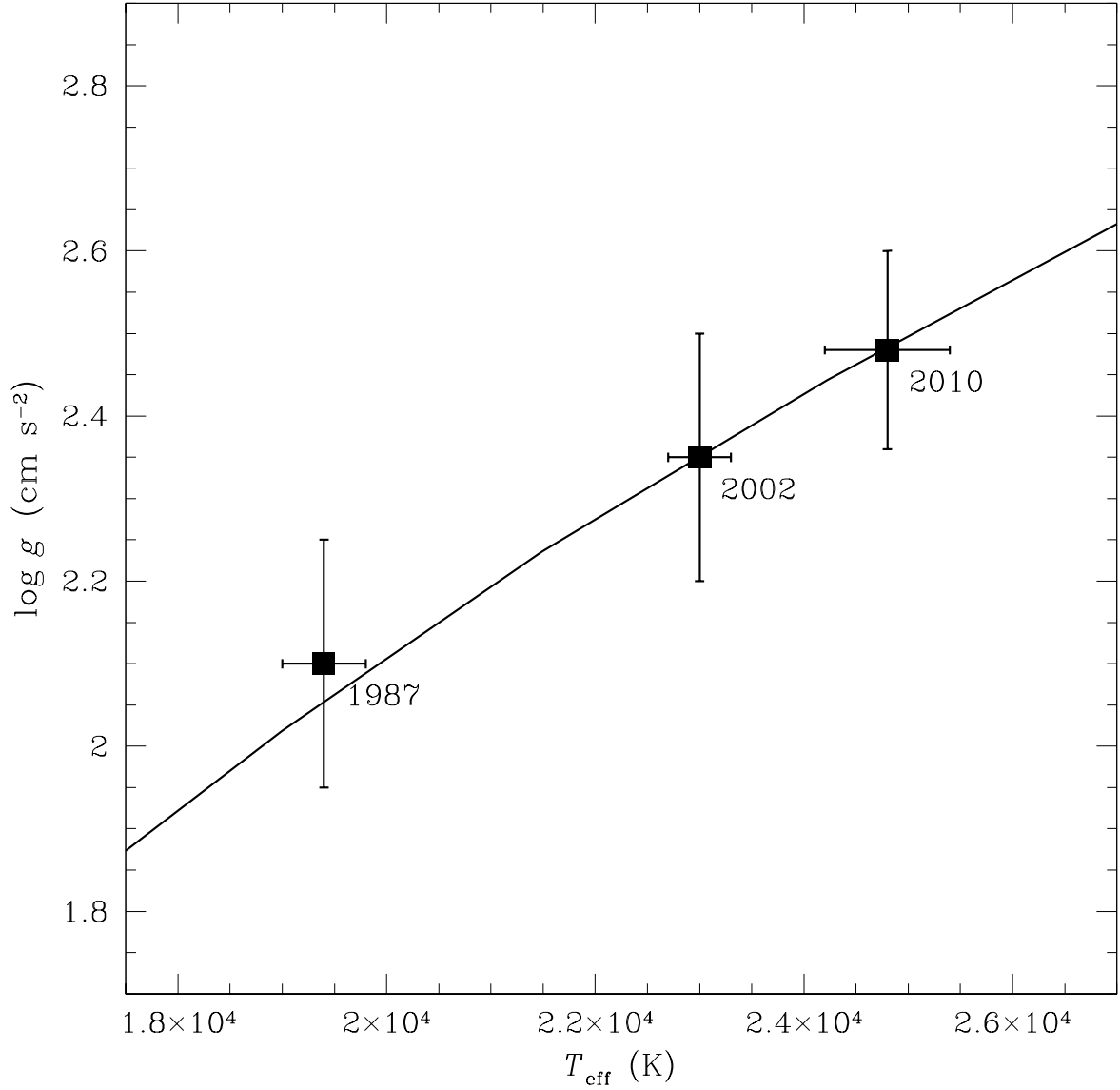}
\caption{The derived stellar parameters, with error bars, for all three epochs.
Evolutionary track from Figure 6 of \citet{saio02} for the merger product corresponding
to a 0.9$M_{\odot}$ model starting from 0.6 - 0.5$M_{\odot}$ CO WDs.
\label{f:gt.time}}
\end{figure}

\begin{figure}
\epsscale{1.00}
\plotone{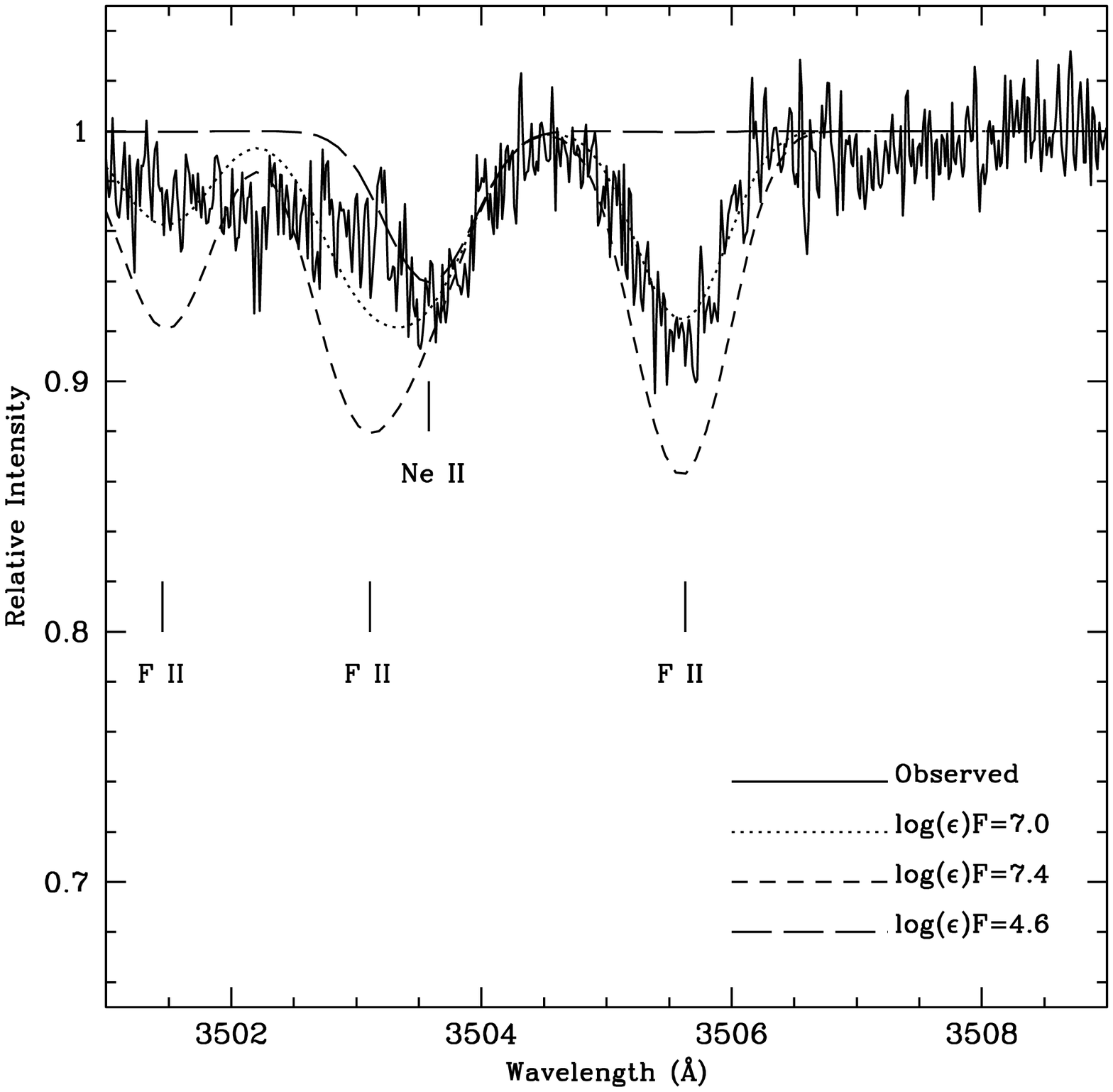}
\caption{The observed UVES spectrum and theoretical F\,{\sc ii} line
profiles calculated using the
LTE model $T_{\rm eff}$=24,750 K and $\log g$ = 2.65, for three different
F abundances $-$ see key on the figure. \label{f:h.nlte}}
\end{figure}

\begin{table}
\caption{Summary of atmospheric parameters}
\label{t:equil}
\begin{tabular}{lcccc}
\hline
Year & $T_{\rm eff}$     & $\log g$       & $\xi$               & $v \sin i$               \\
     &               (K) &          (cgs units) &       (km s$^{-1}$) &            (km s$^{-1}$) \\
\hline
%
%
1987 & $19400\pm400$ &  $2.10\pm0.15$ & $20\pm3$ & 20 $-$ 25  \\
2002 & $23000\pm300$ & $2.35\pm0.15$ & $23\pm3$ & 20 $-$ 27  \\
2010 & $24800\pm600$ & $2.50\pm0.12$ & $24\pm3$ & $40\pm5$  \\

\hline
\end{tabular}
\end{table}

\begin{deluxetable}{lccccc}
\label{t:lines.ucles}
\tabletypesize{\scriptsize}
\tablewidth{0pt}           
\tablecolumns{7}           
\tablecaption{Measured equivalent widths ($W_{\lambda}$) and NLTE photospheric line abundances
for DY\,Cen CASPEC and UCLES spectra.}
\tablehead{                                                                                              
\colhead{} & \colhead{$\chi$} & \colhead{} & \colhead{$W_{\lambda}$: CASPEC, AAT/UCLES} &                
\multicolumn{2}{c}{log $\epsilon(\rm X)$}\\                                                              
\cline{5-6} \\                                                                                           
\colhead{Line} & \colhead{(eV)} & \colhead{log $gf$} & \colhead{(m\AA)} & \colhead{CASPEC\tablenotemark{a}} &
\colhead{AAT/UCLES\tablenotemark{b}}}                                                                        
\startdata                                                                                                   
H\,{\sc i} $\lambda 4101.734$ & 10.199 & $-$0.753 & Synth &10.75 & 10.52     \\
H\,{\sc i} $\lambda 4340.462$ & 10.199 & $-$0.447 & Synth &10.72 & 10.63     \\
H\,{\sc i} $\lambda 4861.323$ & 10.199 & $-$0.020 & Synth &10.58 & 10.70     \\
        &         &         &         &               &               \\
Mean... & \nodata & \nodata & \nodata & 10.70$\pm$0.10 & 10.60$\pm$0.10 \\
        &         &         &         &               &               \\

C\,{\sc ii} $\lambda 4017.272$ & 22.899 & $-$1.031 & 85,\nodata & 9.65 & \nodata    \\                       
C\,{\sc ii} $\lambda 4021.166$ & 22.899 & $-$1.333 & 35, 40 & 9.46 & 9.64    \\                              
C\,{\sc ii} $\lambda 4321.657$ & 23.116 & $-$0.901 & \nodata, 56 & \nodata & 9.45    \\                      
C\,{\sc ii} $\lambda 4323.107$ & 23.114 & $-$1.105 & 49, 32 & 9.49 & 9.38    \\                              
        &         &         &         &               &               \\                                     
C\,{\sc ii} $\lambda 4325.832$ & 23.119 & $-$0.373 &         &         &            \\                       
C\,{\sc ii} $\lambda 4326.164$ & 23.116 & $-$0.407 & 234, 184 &  9.48   & 9.37       \\                      
        &         &         &         &               &               \\                                     
C\,{\sc ii} $\lambda 4372.375$ & 24.656 & $+$0.057\tablenotemark{c} &         &         &            \\      
C\,{\sc ii} $\lambda 4372.501$ & 24.658 & $+$0.272\tablenotemark{c} & 244,\nodata &  9.56   & \nodata    \\  
        &         &         &         &               &               \\                                     
C\,{\sc ii} $\lambda 4374.281$ & 24.654 & $+$0.660\tablenotemark{d} & 302, 225 &  9.69   & 9.32       \\     
C\,{\sc ii} $\lambda 4376.582$ & 24.656 & $+$0.380\tablenotemark{d} & 184,\nodata &  9.38   & \nodata       \\
        &         &         &         &               &               \\                                      
Mean... & \nodata & \nodata & \nodata & 9.53$\pm$0.11 & 9.43$\pm$0.13 \\                                      
        &         &         &         &               &               \\                                      
C\,{\sc iii} $\lambda 4162.877$ & 40.064 & $+$0.218 & \nodata, 85 &  \nodata   & 9.52       \\                
C\,{\sc iii} $\lambda 4186.900$ & 40.010 & $+$0.918 & 52, 150 &  9.63   & 9.43       \\                       
C\,{\sc iii} $\lambda 4647.418$ & 29.535 & $+$0.070 & 261, 483 &  9.40   & 9.19       \\                      
C\,{\sc iii} $\lambda 4651.473$ & 29.535 & $-$0.629 & 153,\nodata &  9.62   & \nodata    \\                   
C\,{\sc iii} $\lambda 4665.860$ & 38.226 & $+$0.044 & \nodata, 90 &  \nodata   & 9.43       \\                
        &         &         &         &               &               \\                                      
Mean... & \nodata & \nodata & \nodata & 9.55$\pm$0.13 & 9.39$\pm$0.14 \\                                      
        &         &         &         &               &               \\                                      
N\,{\sc ii} $\lambda 4227.736$ & 21.599 & $-$0.061 & 74, 61 &  7.68   & 7.67       \\                         
N\,{\sc ii} $\lambda 4447.030$ & 20.409 & $+$0.221 & 216, 150 &  7.86   & 7.62       \\                       
N\,{\sc ii} $\lambda 4601.478$ & 18.466 & $-$0.452 & 192, 241 &  7.82   & 8.18       \\                       
N\,{\sc ii} $\lambda 4607.153$ & 18.462 & $-$0.522 & 174, 189 &  7.80   & 8.03       \\                       
N\,{\sc ii} $\lambda 4613.868$ & 18.466 & $-$0.692 & \nodata, 163  &  \nodata  & 8.08       \\                
N\,{\sc ii} $\lambda 4630.539$ & 18.483 & $+$0.080 & 288, 300 &  7.74   & 7.92       \\                       
N\,{\sc ii} $\lambda 4643.086$ & 18.483 & $-$0.371 & 113, 198 &  7.34?  & 7.92       \\                       
N\,{\sc ii} $\lambda 4788.138$ & 20.654 & $-$0.366 & 46, 48   &  7.47  & 7.62       \\                        
        &         &         &         &               &               \\                                      
Mean... & \nodata & \nodata & \nodata & 7.67$\pm$0.20 & 7.88$\pm$0.21 \\                                      
        &         &         &         &               &               \\                                      
N\,{\sc iii} $\lambda 4200.070$ & 36.856 & $+$0.250 & 12, 30   &  8.24? & 7.85       \\                       
        &         &         &         &               &               \\                                      
O\,{\sc ii} $\lambda 4104.724$ & 25.837 & $-$0.302 &          &        &            \\                        
O\,{\sc ii} $\lambda 4104.990$ & 25.837 & $-$0.015 & 180, 242  &  8.84  & 9.33       \\                       
        &         &         &         &               &               \\                                      
O\,{\sc ii} $\lambda 4153.298$ & 25.837 & $+$0.053 & 176,\nodata  &  8.95  & \nodata      \\                  
O\,{\sc ii} $\lambda 4317.139$ & 22.966 & $-$0.386 & 262, 344  &  8.79  & 8.85       \\                       
O\,{\sc ii} $\lambda 4319.630$ & 22.979 & $-$0.380 & 241, 357  &  8.71  & 8.88       \\                       
O\,{\sc ii} $\lambda 4336.859$ & 22.979 & $-$0.763 & 207, 240  &  8.99  & 8.93       \\                       
O\,{\sc ii} $\lambda 4345.560$ & 22.979 & $-$0.346 & 323, 376  &  8.97  & 8.92       \\                       
O\,{\sc ii} $\lambda 4349.426$ & 22.999 & $+$0.060 & 384, 550  &  8.74  & 8.94       \\                       
O\,{\sc ii} $\lambda 4366.895$ & 22.999 & $-$0.348 & 291, 396  &  8.85  & 8.96       \\                       
O\,{\sc ii} $\lambda 4414.899$ & 23.442 & $+$0.172 & 330, 514  &  8.47  & 8.62       \\                       
O\,{\sc ii} $\lambda 4416.975$ & 23.419 & $-$0.077 & 281, 418  &  8.56  & 8.63       \\                       
O\,{\sc ii} $\lambda 4452.378$ & 23.442 & $-$0.788 & 152, 209  &  8.76  & 8.72       \\                       
O\,{\sc ii} $\lambda 4661.632$ & 22.979 & $-$0.278 & 225, 373  &  8.56  & 8.79       \\                       
O\,{\sc ii} $\lambda 4673.733$ & 22.979 & $-$1.090 & 153, 245  &  9.04  & 9.19       \\                       
O\,{\sc ii} $\lambda 4676.235$ & 22.999 & $-$0.394 & 244, 358  &  8.75  & 8.86       \\                       
O\,{\sc ii} $\lambda 4696.353$ & 22.999 & $-$1.380 &  37, 95   &  8.41  & 8.80       \\                       
O\,{\sc ii} $\lambda 4705.346$ & 26.249 & $+$0.477 & 179,\nodata   &  8.16  & \nodata       \\                
O\,{\sc ii} $\lambda 4906.830$ & 26.305 & $-$0.161 & 112, 235  &  8.47  & 8.53      \\                        
O\,{\sc ii} $\lambda 4924.529$ & 26.305 & $+$0.074 & 175, 300  &  8.53  & 8.45      \\                        
O\,{\sc ii} $\lambda 4941.072$ & 26.554 & $-$0.053 & 88,\nodata   &  9.07  & \nodata       \\                 
O\,{\sc ii} $\lambda 4943.005$ & 26.561 & $+$0.239 & 152,\nodata   &  9.25  & \nodata       \\                
        &         &         &         &               &               \\                                      
Mean... & \nodata & \nodata & \nodata & 8.74$\pm$0.27 & 8.84$\pm$0.22 \\                                      
\enddata                                                                                                      
\tablenotetext{a}{($T_{\rm eff}$, $\log g$, $\xi$)=(19400, 2.10, 20.0)}                                       
\tablenotetext{b}{($T_{\rm eff}$, $\log g$, $\xi$)=(23000, 2.35, 23.0)}                                       
\tablenotetext{c}{\citet{yan87}}                                                                              
\tablenotetext{d}{\citet{wiese66}}                                                                            
\end{deluxetable}

\begin{deluxetable}{lccrcc}
\label{t:lines.uves}
\tabletypesize{\scriptsize}
\tablewidth{0pt}           
\tablecolumns{7}           
\tablecaption{Measured equivalent widths ($W_{\lambda}$) and NLTE/LTE photospheric line abundances
for the DY\,Cen UVES spectrum.}
\tablehead{                                                                               
\colhead{} & \colhead{$\chi$} & \colhead{} & \colhead{$W_{\lambda}$} &                    
\multicolumn{2}{c}{log $\epsilon(\rm X)$}\\                                               
\cline{5-6} \\                                                                            
\colhead{Line} & \colhead{(eV)} & \colhead{log $gf$} & \colhead{(m\AA)} & \colhead{NLTE\tablenotemark{a}} &
\colhead{LTE\tablenotemark{b}}}                                                                            
\startdata                                                                                                 
H\,{\sc i} $\lambda 3721.939$ & 10.199 & $-$1.975 & Synth &10.52 &      \\                                 
H\,{\sc i} $\lambda 3970.072$ & 10.199 & $-$0.993 & Synth &10.72 &      \\                                 
H\,{\sc i} $\lambda 4340.462$ & 10.199 & $-$0.447 & Synth &10.72 &      \\                                 
        &         &         &         &               &               \\                                   
Mean... & \nodata & \nodata & \nodata & 10.65$\pm$0.12 & 10.76$\pm$0.04 \\                                 
        &         &         &         &               &               \\                                   
C\,{\sc ii} $\lambda 3581.757$ & 22.529 & $-$1.643 & 36  &10.10 & 10.15     \\                             
C\,{\sc ii} $\lambda 3952.057$ & 24.278 & $-$0.210\tablenotemark{c} & 152 & 9.82 & 9.85 \\                 
C\,{\sc ii} $\lambda 3977.250$ & 24.372 & $-$0.500\tablenotemark{c} & 66  & 9.58 & 9.62 \\                 
C\,{\sc ii} $\lambda 3980.317$ & 24.373 & $-$0.210\tablenotemark{c} & 105 & 9.57 & 9.60 \\                 
C\,{\sc ii} $\lambda 4017.272$ & 22.899 & $-$1.031\tablenotemark{c} & 74  & 9.83 & 9.90 \\                 
C\,{\sc ii} $\lambda 4021.166$ & 22.899 & $-$1.333\tablenotemark{c} & 39  & 9.79 & 9.86 \\                 
C\,{\sc ii} $\lambda 4313.106$ & 23.116 & $-$0.373 & 202 & 9.96 & 10.01 \\                                 
C\,{\sc ii} $\lambda 4321.657$ & 23.116 & $-$0.901 & 55  & 9.59 & 9.65 \\                                  
C\,{\sc ii} $\lambda 4323.107$ & 23.114 & $-$1.105 & 39  & 9.62 & 9.68 \\                                  
C\,{\sc ii} $\lambda 4374.281$ & 24.654 & $+$0.660 & 353 &10.07 & 10.01 \\                                 
        &         &         &         &               &               \\                                   
Mean... & \nodata & \nodata & \nodata & 9.79$\pm$0.20 & 9.83$\pm$0.19 \\                                   
        &         &         &         &               &               \\                                   
C\,{\sc iii} $\lambda 3885.938$ & 39.852 &$+$0.205 &     &      &      \\                                  
C\,{\sc iii} $\lambda 3886.145$ & 39.852 &$-$0.698 & 152 & 9.62 & 9.77 \\                                  
        &         &         &         &               &               \\
C\,{\sc iii} $\lambda 4056.061$ & 40.197 &$+$0.265 & 160 & 9.66 & 9.72 \\                                  
C\,{\sc iii} $\lambda 4162.877$ & 40.064 &$+$0.218 & 173 & 9.81 & 9.92 \\                                  
C\,{\sc iii} $\lambda 4186.900$ & 40.010 &$+$0.918 & 246 & 9.50 & 9.68 \\                                  
        &         &         &         &               &               \\
C\,{\sc iii} $\lambda 4382.897$ & 39.852 &$-$0.778 &     &      &      \\                                  
C\,{\sc iii} $\lambda 4383.533$ & 39.852 &$-$1.255 & 66  & 9.67 & 9.65 \\                                  
        &         &         &         &               &               \\
C\,{\sc iii} $\lambda 5826.420$ & 40.197 &$+$0.416 & 160  & 9.70 & 9.93 \\                                  
C\,{\sc iii} $\lambda 6350.770$ & 40.197 &$-$0.882 & 17   & 9.83 & 9.80 \\                                  
        &         &         &         &               &               \\                                   
Mean... & \nodata & \nodata & \nodata & 9.68$\pm$0.11 & 9.78$\pm$0.11 \\                                   
        &         &         &         &               &               \\                                   
C\,{\sc iv} $\lambda 5801.310$ & 37.548 &$-$0.194 & 42: & 9.72 & 9.80 \\                                   
C\,{\sc iv} $\lambda 5811.970$ & 37.548 &$-$0.495 & 24: & 9.86 & 9.60 \\                                   
        &         &         &         &               &               \\                                   
Mean... & \nodata & \nodata & \nodata & 9.79$\pm$0.10 & 9.70$\pm$0.14 \\                                   
        &         &         &         &               &               \\                                   
N\,{\sc ii} $\lambda 3437.145$ & 18.497 & $-$0.436 & 112 &  7.97  & 8.13 \\                                
N\,{\sc ii} $\lambda 3955.851$ & 18.466 & $-$0.813 &  78 &  8.01  & 8.22 \\                                
N\,{\sc ii} $\lambda 3994.997$ & 18.497 & $+$0.208 & 349 &  8.32? & 8.42 \\                                
N\,{\sc ii} $\lambda 4035.081$ & 23.124 & $+$0.623\tablenotemark{c} &  85 &  7.75  & 7.87 \\               
N\,{\sc ii} $\lambda 4041.310$ & 23.142 & $+$0.853\tablenotemark{c} & 135 &  7.82  & 7.93 \\               
N\,{\sc ii} $\lambda 4043.532$ & 23.132 & $+$0.743\tablenotemark{c} &  78 &  7.58  & 7.70 \\               
N\,{\sc ii} $\lambda 4171.595$ & 23.196 & $+$0.280\tablenotemark{c} &  35 &  7.64  & 7.77 \\               
N\,{\sc ii} $\lambda 4176.159$ & 23.196 & $+$0.600\tablenotemark{c} &  65 &  7.64  & 7.76 \\               
N\,{\sc ii} $\lambda 4227.736$ & 21.600 & $-$0.061 &  50 &  7.78  & 7.93 \\                                
        &         &         &         &               &               \\                                   
N\,{\sc ii} $\lambda 4241.755$ & 23.242 & $+$0.210\tablenotemark{c} &     &        &      \\               
N\,{\sc ii} $\lambda 4241.786$ & 23.246 & $+$0.713\tablenotemark{c} & 103 &  7.69  & 7.92 \\               
        &         &         &         &               &               \\                                   
        &         &         &         &               &               \\                                   
N\,{\sc ii} $\lambda 4432.736$ & 23.415 & $+$0.580\tablenotemark{c} &  80 &  7.83  & 7.95 \\               
        &         &         &         &               &               \\                                   
Mean... & \nodata & \nodata & \nodata & 7.77$\pm$0.14 & 7.92$\pm$0.16 \\                                   
        &         &         &         &               &               \\                                   
N\,{\sc iii} $\lambda 4200.070$ & 36.856 & $+$0.250 &  42 &  7.78  & 8.42 \\                               
        &         &         &         &               &               \\                                   
O\,{\sc ii} $\lambda 3305.003$ & 25.849 & $-$0.723  & 85  & 9.55? & 9.16 \\                                
O\,{\sc ii} $\lambda 3306.451$ & 25.837 & $-$0.740  & 80  & 9.53? & 9.14 \\                                
O\,{\sc ii} $\lambda 3377.146$ & 25.286 & $-$0.342  & 168 & 8.89  & 9.15 \\                                
O\,{\sc ii} $\lambda 3390.209$ & 25.286 & $-$0.044  & 212 & 8.92  & 9.07 \\                                
        &         &         &         &               &               \\                                   
O\,{\sc ii} $\lambda 3407.223$ & 28.510 & $-$1.121  &     &       &      \\                                
O\,{\sc ii} $\lambda 3407.276$ & 28.512 & $+$0.025  & 130 & 9.37  & 9.30 \\                                
        &         &         &         &               &               \\                                   
O\,{\sc ii} $\lambda 3409.706$ & 28.512 & $-$0.167  &     &       &      \\                                
O\,{\sc ii} $\lambda 3409.760$ & 28.512 & $-$1.121  & 65  & 9.03  & 9.02 \\                                
        &         &         &         &               &               \\                                   
O\,{\sc ii} $\lambda 3739.761$ & 26.305 & $-$0.427  & 175 & 8.85  & 9.19 \\                                
O\,{\sc ii} $\lambda 3762.465$ & 26.305 & $-$0.580  & 154 & 8.92  & 9.23 \\                                
O\,{\sc ii} $\lambda 3882.194$ & 25.664 & $-$0.035  & 290 & 9.48? & 9.15 \\                                
O\,{\sc ii} $\lambda 3893.518$ & 25.638 & $-$1.589  & 32  & 9.19  &      \\                                
O\,{\sc ii} $\lambda 3907.455$ & 25.649 & $-$0.925  & 105 & 9.39  & 9.11 \\                                
        &         &         &         &               &               \\                                   
O\,{\sc ii} $\lambda 3911.958$ & 25.661 & $+$0.000  &     &       &      \\                                
O\,{\sc ii} $\lambda 3912.117$ & 25.661 & $-$0.888  & 342 & 9.52? & 9.16 \\                                
        &         &         &         &               &               \\                                   
O\,{\sc ii} $\lambda 3945.038$ & 23.419 & $-$0.727  & 330 & 8.91  & 9.47?\\                                
O\,{\sc ii} $\lambda 3954.362$ & 23.419 & $-$0.396  & 423 & 8.80  & 9.51?\\                                
O\,{\sc ii} $\lambda 3973.256$ & 23.441 & $-$0.015  & 585 & 8.83  &      \\                                
O\,{\sc ii} $\lambda 3982.714$ & 23.441 & $-$0.703  & 340 & 8.90  & 9.50?\\                                
O\,{\sc ii} $\lambda 4185.440$ & 28.357 & $+$0.604  & 240 & 8.96  & 8.84 \\                                
        &         &         &         &               &               \\                                   
O\,{\sc ii} $\lambda 4189.581$ & 28.360 & $-$0.828  &     &       &      \\                                
O\,{\sc ii} $\lambda 4189.794$ & 28.360 & $+$0.716  & 253 & 8.89  & 8.77 \\                                
        &         &         &         &               &               \\                                   
O\,{\sc ii} $\lambda 4192.512$ & 28.509 & $-$0.470  & 48  & 8.83  & 8.89 \\                                
        &         &         &         &               &               \\                                   
O\,{\sc ii} $\lambda 4196.273$ & 28.512 & $-$1.425  &     &       &      \\                                
O\,{\sc ii} $\lambda 4196.697$ & 28.512 & $-$0.726  & 63  & 9.13  & 9.20 \\                                
        &         &         &         &               &               \\                                   
O\,{\sc ii} $\lambda 4327.460$ & 28.509 & $+$0.057  &     &       &      \\                                
O\,{\sc ii} $\lambda 4327.849$ & 28.509 & $-$1.090  & 140 & 8.93  & 8.93 \\                                
        &         &         &         &               &               \\                                   
O\,{\sc ii} $\lambda 4336.859$ & 22.979 & $-$0.763  & 345 & 9.16  & 9.58?\\                                
O\,{\sc ii} $\lambda 4345.560$ & 22.979 & $-$0.346  & 465 & 9.20  &      \\                                
O\,{\sc ii} $\lambda 4349.426$ & 22.998 & $+$0.060  & 687 & 9.33  &      \\                                
O\,{\sc ii} $\lambda 4366.895$ & 22.999 & $-$0.348  & 570 & 9.48? &      \\                                
O\,{\sc ii} $\lambda 4395.935$ & 26.248 & $-$0.167  & 190 & 8.93  & 8.99 \\                                
O\,{\sc ii} $\lambda 4405.978$ & 26.248 & $-$1.300  & 29  & 8.98  & 8.99 \\                                
O\,{\sc ii} $\lambda 4414.899$ & 23.441 & $+$0.172  & 665 & 9.10  &      \\                                
O\,{\sc ii} $\lambda 4416.975$ & 23.419 & $-$0.077  & 530 & 9.03  &      \\                                
O\,{\sc ii} $\lambda 4452.378$ & 23.442 & $-$0.788  & 265 & 9.04  & 9.30 \\                                
O\,{\sc ii} $\lambda 4443.010$ & 28.358 & $-$0.047  & 156 & 9.17  & 9.13 \\                                
O\,{\sc ii} $\lambda 4448.191$ & 28.361 & $+$0.083  & 188 & 9.22  & 9.17 \\                                
O\,{\sc ii} $\lambda 6641.031$ & 23.419 & $-$0.884  & 394 & 9.55? &      \\                                
O\,{\sc ii} $\lambda 6721.388$ & 23.441 & $-$0.610  & 582 & 9.64? &      \\                                
        &         &         &         &               &               \\                                   
Mean... & \nodata & \nodata & \nodata & 9.03$\pm$0.17 & 9.09$\pm$0.14 \\                                   
        &         &         &         &               &               \\                                   
O\,{\sc iii} $\lambda 3312.329$ & 33.151 & $-$0.644  & 36  & 8.93  & 9.26 \\                               
O\,{\sc iii} $\lambda 3340.765$ & 33.182 & $-$0.482  & 56  & 9.04  & 9.47 \\                               
O\,{\sc iii} $\lambda 3430.568$ & 37.250 & $-$0.902  & 18  & 9.15  & 9.32 \\                               
O\,{\sc iii} $\lambda 3444.052$ & 37.250 & $-$0.427  & 34  & 9.02  & 9.29 \\                               
O\,{\sc iii} $\lambda 3707.272$ & 37.234 & $-$0.121  & 31  & 8.70  & 8.83 \\                               
O\,{\sc iii} $\lambda 3715.086$ & 37.249 & $+$0.149  & 56  & 8.82  & 9.04 \\                               
O\,{\sc iii} $\lambda 3754.696$ & 33.150 & $-$0.099  & 126 & 8.83  & 9.26 \\                               
O\,{\sc iii} $\lambda 3757.232$ & 33.135 & $-$0.452  & 86  & 8.88  & 9.16 \\                               
O\,{\sc iii} $\lambda 3759.875$ & 33.182 & $+$0.162  & 157 & 8.73  & 9.22 \\                               
O\,{\sc iii} $\lambda 3774.026$ & 33.150 & $-$0.601  & 60  & 8.81  & 9.08 \\                               
O\,{\sc iii} $\lambda 3810.985$ & 33.182 & $-$1.810  & 8   & 8.79  & 8.88 \\                               
O\,{\sc iii} $\lambda 3961.573$ & 38.011 & $+$0.314  & 60  & 9.05  & 9.04 \\                               
        &         &         &         &               &               \\                                   
Mean... & \nodata & \nodata & \nodata & 8.90$\pm$0.14 & 9.15$\pm$0.19 \\                                   
\enddata
\tablenotetext{a}{($T_{\rm eff}$, $\log g$, $\xi$)=(25000, 2.50, 24.0)}
\tablenotetext{b}{($T_{\rm eff}$, $\log g$, $\xi$)=(24750, 2.65, 30.0)}
\tablenotetext{c}{Kurucz $gf$-value}
\end{deluxetable}

\begin{deluxetable}{lccrcc}
\label{t:lines.uvesh}
\tabletypesize{\scriptsize}
\tablewidth{0pt}
\tablecolumns{7}
\tablecaption{Measured equivalent widths ($W_{\lambda}$) and NLTE/LTE photospheric line abundances
of F, Ne and heavier elements for the DY\,Cen UVES spectrum.}
\tablehead{
\colhead{} & \colhead{$\chi$} & \colhead{} & \colhead{$W_{\lambda}$} &
\multicolumn{2}{c}{log $\epsilon(\rm X)$}\\
\cline{5-6} \\
\colhead{Line} & \colhead{(eV)} & \colhead{log $gf$} & \colhead{(m\AA)} & \colhead{NLTE\tablenotemark{a}} &
\colhead{LTE\tablenotemark{b}}}
\startdata
F\,{\sc ii} $\lambda 3502.840$ & 25.102 & $-$0.400 &     &       &       \\                               
F\,{\sc ii} $\lambda 3502.964$ & 25.102 & $+$0.187 &     &       &       \\                               
F\,{\sc ii} $\lambda 3503.109$ & 25.102 & $+$0.391 & 42  &       & 6.90  \\                               
        &         &         &         &               &               \\
F\,{\sc ii} $\lambda 3505.368$ & 25.104 & $-$0.757 &     &       &       \\                               
F\,{\sc ii} $\lambda 3505.513$ & 25.104 & $+$0.090 &     &       &       \\                               
F\,{\sc ii} $\lambda 3505.628$ & 25.104 & $+$0.676 & 70  &       & 7.05  \\                               
        &         &         &         &               &               \\
Mean... & \nodata & \nodata & \nodata &               & 6.98$\pm$0.11 \\
        &         &         &         &               &               \\
Ne\,{\sc i} $\lambda 6143.063$ & 16.619 & $-$0.100 & 141 & 8.76  & 9.41  \\                               
Ne\,{\sc i} $\lambda 6163.594$ & 16.715 & $-$0.620 & 70  & 8.70  & 9.54  \\                               
Ne\,{\sc i} $\lambda 6266.495$ & 16.715 & $-$0.370 & 84  & 8.53  & 9.39  \\                               
Ne\,{\sc i} $\lambda 6334.428$ & 16.619 & $-$0.320 & 115 & 8.87  & 9.50  \\                               
Ne\,{\sc i} $\lambda 6402.246$ & 16.619 & $+$0.330 & 245 & 8.63  & 9.44  \\      
        &         &         &         &               &               \\                                   
Mean... & \nodata & \nodata & \nodata & 8.70$\pm$0.13 & 9.46$\pm$0.06 \\                                   
        &         &         &         &               &               \\                                   
Ne\,{\sc ii} $\lambda 3309.739$ & 27.783 & $-$0.990 & 30  & 8.09  & 8.36  \\                               
Ne\,{\sc ii} $\lambda 3323.734$ & 27.783 & $+$0.030 & 132 & 7.87  & 8.47  \\                               
Ne\,{\sc ii} $\lambda 3327.152$ & 27.233 & $-$0.220 & 130 & 7.92  & 8.58  \\                               
Ne\,{\sc ii} $\lambda 3334.836$ & 27.169 & $+$0.380 & 235 & 7.72  & 8.61  \\                               
Ne\,{\sc ii} $\lambda 3344.396$ & 27.270 & $-$0.300 & 140 & 8.05  & 8.74  \\                               
Ne\,{\sc ii} $\lambda 3345.454$ & 30.549 & $-$0.030 & 73: & 8.08  & 8.69  \\                               
Ne\,{\sc ii} $\lambda 3357.820$ & 30.927 & $-$0.290 & 37  & 8.15  & 8.59  \\                               
Ne\,{\sc ii} $\lambda 3378.217$ & 27.659 & $-$0.240 & 72  & 7.79  & 8.29  \\                               
Ne\,{\sc ii} $\lambda 3388.419$ & 31.185 & $+$0.360 & 54  & 7.93  & 8.23  \\                               
Ne\,{\sc ii} $\lambda 3417.689$ & 31.121 & $+$0.350 & 66  & 8.22  & 8.37  \\                               
Ne\,{\sc ii} $\lambda 3453.068$ & 31.185 & $-$0.480 & 30  & 8.11  & 8.44  \\                               
Ne\,{\sc ii} $\lambda 3481.933$ & 27.783 & $-$0.290 & 102 & 7.97  & 8.28  \\                               
Ne\,{\sc ii} $\lambda 3542.845$ & 31.362 & $+$0.130 & 82  & 7.93  & 8.48  \\                               
Ne\,{\sc ii} $\lambda 3557.803$ & 27.859 & $-$1.140 & 25  & 8.07  & 8.34  \\                               
Ne\,{\sc ii} $\lambda 3565.826$ & 31.362 & $-$0.330 & 32  & 8.01  & 8.38  \\                               
Ne\,{\sc ii} $\lambda 3568.500$ & 30.549 & $+$0.330 & 136 & 7.88  & 8.51  \\                               
Ne\,{\sc ii} $\lambda 3571.230$ & 31.362 & $-$0.320 & 45  & 8.17  & 8.55  \\                               
Ne\,{\sc ii} $\lambda 3574.611$ & 30.549 & $+$0.170 & 126 & 7.99  & 8.60  \\                               
Ne\,{\sc ii} $\lambda 3628.036$ & 31.512 & $-$0.320 & 20  & 7.95  & 8.18  \\                               
Ne\,{\sc ii} $\lambda 3643.928$ & 27.783 & $-$0.590 & 56  & 7.82  & 8.21  \\                               
Ne\,{\sc ii} $\lambda 3664.073$ & 27.169 & $-$0.250 & 172 & 7.92  & 8.58  \\                               
Ne\,{\sc ii} $\lambda 3694.212$ & 27.169 & $+$0.090 & 233 & 7.75  & 8.51  \\                               
Ne\,{\sc ii} $\lambda 3709.621$ & 27.233 & $-$0.340 & 170 & 7.93  & 8.61  \\                               
Ne\,{\sc ii} $\lambda 3766.258$ & 27.233 & $-$0.430 & 156 & 7.92  & 8.57  \\                               
Ne\,{\sc ii} $\lambda 3777.134$ & 27.270 & $-$0.440 & 190 & 8.06  & 8.76  \\                               
Ne\,{\sc ii} $\lambda 4150.690$ & 34.644 & $-$0.030\tablenotemark{c} & 22  & 8.16  & 8.46  \\                               
Ne\,{\sc ii} $\lambda 4217.169$ & 34.609 & $+$0.090\tablenotemark{c} & 15  & 7.86  & 8.16  \\                               
Ne\,{\sc ii} $\lambda 4219.745$ & 34.609 & $+$0.750\tablenotemark{c} & 96  & 8.17  & 8.57  \\                               
        &         &         &         &               &               \\                                   
Ne\,{\sc ii} $\lambda 4231.532$ & 34.619 & $-$0.080\tablenotemark{c} &     &       &       \\                               
Ne\,{\sc ii} $\lambda 4231.636$ & 34.619 & $+$0.260\tablenotemark{c} & 26  & 7.78  & 8.10  \\                               
        &         &         &         &               &               \\                                   
Ne\,{\sc ii} $\lambda 4239.911$ & 34.632 & $-$0.490\tablenotemark{c} &     &       &       \\                               
Ne\,{\sc ii} $\lambda 4240.105$ & 34.632 & $-$0.020\tablenotemark{c} & 10  &  7.67 & 7.97  \\                               
        &         &         &         &               &               \\                                   
Ne\,{\sc ii} $\lambda 4250.645$ & 34.632 & $+$0.150\tablenotemark{c} & 25  & 8.04  & 8.36  \\                               
Ne\,{\sc ii} $\lambda 4391.991$ & 34.737 & $+$0.920\tablenotemark{c} & 82  & 7.89  & 8.32  \\                               
        &         &         &         &               &               \\                                   
Mean... & \nodata & \nodata & \nodata & 7.96$\pm$0.14 & 8.43$\pm$0.19 \\                                   
        &         &         &         &               &               \\                                   
Mg\,{\sc ii} $\lambda 4481.126$ & 8.863 & $+$0.749  &          &        &      \\                              
Mg\,{\sc ii} $\lambda 4481.150$ & 8.863 & $-$0.553  &          &        &      \\                              
Mg\,{\sc ii} $\lambda 4481.325$ & 8.863 & $+$0.594  & Synth    & 6.76   & 7.13 \\                              
        &         &         &         &               &               \\
Al\,{\sc iii} $\lambda 3601.630$ & 14.376 & $+$0.01  &     &       &      \\                              
Al\,{\sc iii} $\lambda 3601.927$ & 14.374 & $-$0.95  & 100 &       & 6.07 \\                              
        &         &         &         &               &               \\                                   
Al\,{\sc iii} $\lambda 4149.913$ & 20.55  & $+$0.63  & $\leq$8 &       & $\leq$5.06? \\                        
        &         &         &         &               &               \\
Al\,{\sc iii} $\lambda 4479.885$ & 20.78  & $+$0.09  &     &       &       \\
Al\,{\sc iii} $\lambda 4479.971$ & 20.78  & $+$1.02  &     &       &       \\
Al\,{\sc iii} $\lambda 4480.009$ & 20.78  & $-$0.53  & 95  &       & 6.22  \\
        &         &         &         &               &               \\
Al\,{\sc iii} $\lambda 5722.730$ & 15.642 & $-$0.07  & 252 &       & 7.14? \\                              
        &         &         &         &               &               \\                                   
Mean... & \nodata & \nodata & \nodata &               & 6.15$\pm$0.11 \\                                   
        &         &         &         &               &               \\                                   
Si\,{\sc iii} $\lambda 5739.734$ & 19.722 & $-$0.096  & 300  & 7.23  & 7.50  \\                             
        &         &         &         &               &               \\                                   
Si\,{\sc iv} $\lambda 4116.104$ & 24.050 & $-$0.110  & 176  & 6.69  & 7.03 \\                              
        &         &         &         &               &               \\                                   
P\,{\sc iii} $\lambda 4080.089$ & 14.490 & $-$0.310\tablenotemark{d}  & 20   &       & 5.33? \\             
P\,{\sc iii} $\lambda 4222.198$ & 14.610 & $+$0.210\tablenotemark{d}  & 157  &       & 5.91 \\             
        &         &         &         &               &               \\                                   
Mean... & \nodata & \nodata & \nodata &               & 5.91$\pm$0.00 \\                                   
        &         &         &         &               &               \\                                   
P\,{\sc iv} $\lambda 3347.736$ & 28.132 & $+$0.25\tablenotemark{e}   & 19   &       & 6.02 \\
P\,{\sc iv} $\lambda 3364.467$ & 28.132 & $+$0.02\tablenotemark{e}   & 12   &       & 6.01 \\
P\,{\sc iv} $\lambda 4249.656$ & 29.012 & $-$0.13\tablenotemark{e}   & 11   &       & 5.85 \\
        &         &         &         &               &               \\
Mean... & \nodata & \nodata & \nodata &               & 5.96$\pm$0.10 \\
        &         &         &         &               &               \\
S\,{\sc iii} $\lambda 3324.854$ & 17.745 & $+$0.057  & 40  & 6.11  & 6.05 \\              
S\,{\sc iii} $\lambda 3656.560$ & 18.187 & $-$0.921  & 6   & 6.20  & 6.08 \\              
S\,{\sc iii} $\lambda 3661.942$ & 18.192 & $-$0.462  & 15  & 6.12  & 6.06 \\              
S\,{\sc iii} $\lambda 3717.771$ & 18.244 & $-$0.060  & 65  & 6.29  & 6.32 \\              
S\,{\sc iii} $\lambda 3778.903$ & 18.193 & $-$1.148  & 7   & 6.39  & 6.33 \\              
        &         &         &         &               &               \\                                   
Mean... & \nodata & \nodata & \nodata & 6.22$\pm$0.12 & 6.17$\pm$0.14 \\                                   
        &         &         &         &               &               \\                                   
Fe\,{\sc iii} $\lambda 3603.888$ & 11.210 & $-$1.380\tablenotemark{c}  &  13  & 6.00  & 6.61 \\            
Fe\,{\sc iii} $\lambda 4005.039$ & 11.570 & $-$1.760\tablenotemark{c}  &  $\leq$9  & $\leq$6.10  & $\leq$6.83 \\            
Fe\,{\sc iii} $\lambda 4122.780$ & 20.599 & $+$0.360\tablenotemark{c}  &  $\leq$8  & $\leq$6.40  & $\leq$6.75 \\            
Fe\,{\sc iii} $\lambda 4137.764$ & 20.613 & $+$0.630\tablenotemark{c}  &  $\leq$8  & $\leq$6.00  & $\leq$6.47 \\            
Fe\,{\sc iii} $\lambda 4139.350$ & 20.613 & $+$0.520\tablenotemark{c}  &  $\leq$8  & $\leq$6.20  & $\leq$6.57 \\            
Fe\,{\sc iii} $\lambda 4140.482$ & 20.613 & $+$0.100\tablenotemark{c}  &  $\leq$12 & $\leq$6.80  & $\leq$7.18 \\            
Fe\,{\sc iii} $\lambda 4164.731$ & 20.634 & $+$0.920\tablenotemark{c}  &  $\leq$9  & $\leq$5.90  & $\leq$6.24 \\            
Fe\,{\sc iii} $\lambda 4166.840$ & 20.634 & $+$0.410\tablenotemark{c}  &  $\leq$9  & $\leq$6.40  & $\leq$6.75 \\            
        &         &         &         &               &               \\                                   
Mean... & \nodata & \nodata & \nodata & 6.00$\pm$0.00 & 6.61$\pm$0.00 \\                                   
\enddata                                                                                                   
\tablenotetext{a}{($T_{\rm eff}$, $\log g$, $\xi$)=(25000, 2.50, 24.0)}                                    
\tablenotetext{b}{($T_{\rm eff}$, $\log g$, $\xi$)=(24750, 2.65, 30.0)}                                    
\tablenotetext{c}{Kurucz $gf$-value}                                                                       
\tablenotetext{d}{\citet{wiese69}}                                                                         
\tablenotetext{e}{\citet{froese2006}}                                                                      
\end{deluxetable}

\begin{table}
\caption{Summary of photospheric abundances}
\label{t:equil}
\begin{tabular}{lrrrr}
\hline
Element & non-LTE & LTE & JH93 & Sun\tablenotemark{a} \\
\hline
H   & 10.7 & 10.8 & 10.8 & 12.0 \\
He   & 11.5 & 11.5 & 11.5 & 10.9 \\
C   & 9.6 & 9.8 & 9.5 & 8.4 \\
N   & 7.8 & 7.9 & 8.0 & 7.8 \\
O   & 9.0 & 9.1 & 8.9 & 8.7 \\
F   & \nodata & 7.1 & \nodata & 4.6 \\
Ne\,{\sc i}   & 8.7 & 9.5 & 9.6 & 7.9 \\
Ne\,{\sc ii}   & 8.0 & 8.4 & \nodata & \nodata \\
Mg  & 6.7 & 7.1 & 7.3 & 7.6 \\
Al  & \nodata & 6.1 & 5.9 & 6.5 \\
Si  & 6.8 & 7.0 & 8.1 & 7.5 \\
P  & \nodata & 6.0 & 5.8 & 5.4 \\
S  & 6.2 & 6.2 & 7.1 & 7.1 \\
Ar  & \nodata & $<$6.9 & 6.1 & 6.4 \\
Fe  & 6.0 & 6.6 & 5.0 & 7.5 \\
\end{tabular}
\tablenotetext{a}{\citet{asplund09}}
\end{table}

\clearpage


\end{document}